\documentclass[twocolumn]{aastex61}

\newcommand{\cellspace}{\rule{0pt}{2.8ex}}
\accepted{September 25, 2017}

\shorttitle{X/$\gamma$-ray correlation in NGC 4945}
\shortauthors{Wojaczy\'nski \& Nied\'zwiecki}

\begin{document}

\title{The X/$\gamma$-ray correlation in NGC 4945 and the nature of its $\gamma$-ray source}

\email{rafal.wojaczynski@wp.pl, niedzwiecki@uni.lodz.pl}

\author{Rafa{\l} Wojaczy\'nski}
\altaffiliation{Dip. di Fisica, Universita di Trieste, Via Valerio 2, I-34127 Trieste, Italy\\}
\altaffiliation{INFN Trieste, Padriciano 99, I-34012 Trieste, Italy\\}
\affiliation{Department of Astrophysics, University of \L \'od\'z, Pomorska 149/153, 
90-236 \L \'od\'z, Poland\\}

\author{Andrzej Nied\'zwiecki}
\affiliation{Department of Astrophysics, University of \L \'od\'z, Pomorska 149/153, 
90-236 \L \'od\'z, Poland\\}

\begin{abstract}
We report hints for the correlation between the X-ray and $\gamma$-ray emission in the nearby galaxy NGC 4945, which harbors both an active galactic nucleus and a nuclear starburst region. We have divided the {\it Fermi}/LAT observations of NGC 4945 into two datasets, comprising events detected during the low (L) and high (H) level of X-ray emission from the active nucleus of this galaxy, determined using the {\it Swift}/BAT light curve. Both datasets contain an equal amount of 3.8 years of LAT data and NGC 4945 is detected with a similar statistical significance of $\sim 15 \sigma$ in L and $14 \sigma$ in H. However, the slope of the $\gamma$-ray spectrum hardens with the increase of the X-ray flux, with the photon index $\Gamma = 2.47 \pm 0.07$ in L and $2.11 \pm 0.08$ in H. The change is confirmed by a systematic variation of the spectral energy distribution as well as a substantial reversal of the $\gamma$-ray signal in significance maps for low and high $\gamma$-ray energies. The X/$\gamma$-ray correlation indicates that the $\gamma$-ray production is dominated by the active nucleus rather than by cosmic rays interacting with the interstellar medium. We discuss possible locations of the  $\gamma$-ray source. We also compare NGC 4945 with other starburst galaxies detected by LAT and we note similarities between those with active nuclei, e.g.\ unlikely high  efficiencies of $\gamma$-ray production in starburst scenario, which argues for a significant contribution of their active nuclei to the $\gamma$-ray emission. 
\end{abstract}


\section{Introduction}

NGC 4945 is one of the nearest AGNs (D = 3.8 Mpc), with the black hole mass of $M=1.4  \times  10^6 M_{\odot}$ from megamaser measurements \citep{1997ApJ...481L..23G}. It is the brightest Seyfert 2 galaxy in the hard X-ray range, radiating at a variable rate of $L/L_\mathrm{Edd} \sim 0.1$ \citep{2000ApJ...535L..87M}. Its X-ray spectrum shows a strong photoelectric absorption, with a column density $N_{\rm H} \simeq 4 \times 10^{24}$ cm$^{-2}$ \citep{2003ApJ...588..763D}, at which its nucleus can be directly seen above $\sim 8$ keV. The observed hard X-ray radiation is highly variable, by a factor of several on a time scale of days, confirming that it is a transmission-dominated Compton-thick AGN.

NGC 4945 is also one of a few radio-quiet AGNs detected by {\it Fermi}/LAT \citep{2010ApJS..188..405A,2010A&A...524A..72L}. The origin of this $\gamma$-ray signal is unclear, as this galaxy hosts a circumnuclear sturburst \citep[e.g.][]{2009AJ....137..537L} which may also account for this emission \citep[e.g.][]{2016CRPhy..17..585O}. Variability studies are crucial to disentangle the role of the AGN and starburst activities, but overcoming the weakness of the $\gamma$-ray signal is a major issue for such studies. The apparent lack of the $\gamma$-ray variability, assessed in \citet{2012ApJ...755..164A},  could favor the $\gamma$-ray production dominated by starburst processes. However, the 3-month intervals used in \citet{2012ApJ...755..164A} are too short to accumulate a statistically significant $\gamma$-ray signal from this galaxy, whereas its AGN exhibits  an approximately constant activity on such a time-scale.

In this work we revisit the issue of $\gamma$-ray variability in NGC 4945 and we search for a correlation between its X-ray and $\gamma$-ray emission. We did not attempt to find it in observations carried out  continuously over the time sufficient for an adequate significance of the LAT measurement (i.e.\ at least a year), as this could only probe the averaged out X-ray as well as $\gamma$-ray emission, even if the latter followed the changes of the former. Instead, we considered intermittent data sets including LAT data corresponding to different X-ray flux levels. This allowed us to reveal  the change of the $\gamma$-ray spectrum related with the change of the X-ray flux. The implied constraints on the $\gamma$-ray source, and comparison with other radio-quiet galaxies detected in $\gamma$-rays, are discussed in Section \ref{sect:disc}

\section{Observational data}
\label{sec:data}

We use the {\it Fermi}/LAT and {\it Swift}/BAT data from observations performed by these detectors between 2008 August 4 and 2016 August 15. NGC 4945 exhibits hard X-ray flux variations of a factor of two on a timescale of $2 \times 10^4$ s and of a factor of five on a timescale of several days   \citep{2000ApJ...535L..87M,2014ApJ...793...26P}. Then, the daily count rate values, $\mathcal F_\mathrm{X}$, from the BAT survey program \citep{2013ApJS..209...14K} are convenient to probe changes in this source. Using them we divide all days with contemporaneous LAT and BAT measurements into MJD sets comprising days with various $\mathcal F_\mathrm{X}$ ranges, which then allows us to study the $\gamma$-ray spectral parameters corresponding to different activity levels of the AGN.

\subsection{BAT}
\label{sec:BAT}

The good  quality data (with {\tt DATA\_FLAG=0}) from BAT light curves\footnote{http://swift.gsfc.nasa.gov/results/transients} in the 15--50 keV range  allow to determine $\mathcal F_\mathrm{X}$ for 2783 days out of 2821 days in the considered period of time. We split them into two approximately equal MJD sets, containing days with $\mathcal F_\mathrm{X}$ lower (set L; 1393 days)  and higher (set H; 1390 days) than $1.71 \times 10^{-3}$ cts cm$^{-2}$ s$^{-1}$. To test the effect of possible misclassification of the X-ray flux level due to a short exposure time, we also define sets denoted by L5,H5 and L10,H10, using a similar procedure but only for days with the total exposure time of at least 5 and 10 ksec, respectively. Here, with the median rate of $\mathcal F_\mathrm{X} =1.67   \times   10^{-3}$ cts cm$^{-2}$ s$^{-1}$, we get 920 days in both L5 and H5, and with $\mathcal F_\mathrm{X} =1.69 \times   10^{-3}$ cts cm$^{-2}$ s$^{-1}$, we get 418 days in L10 and 420 days in H10. The average BAT count rate for each set, $\overline{\mathcal F}_\mathrm{X}$, computed using equations A2 in \cite{2008ApJ...673...96A}, is given in Table \ref{tab:PLfit}. $\overline{\mathcal F}_\mathrm{X}$ in L,L5,L10 is lower by a factor of $\sim$5 than in H,H5,H10.

\subsection{LAT}
\label{sec:LAT}

We performed the maximum likelihood analysis using {\tt Pass 8} LAT data in the 0.1--100 GeV range and {\tt v10r0p5} {\it Fermi} Science Tools with the P8R2\_SOURCE\_ V6 instrument response function.  We used the standard templates for the Galactic (\texttt{gll\_iem\_v06.fits}) and the isotropic (\texttt{iso\_P8R2\_SOURCE\_V6\_v06.txt}) backgrounds. 

Our results presented in Section \ref{sec:results} were obtained using unbinned likelihood analysis, for which events were selected from the region with the radius of $10^\circ$ centered on the position of NGC 4945. We took into account all sources reported in the 4-year LAT  Catalog \citep[][hereafter 3FGL]{2015ApJS..218...23A} within the radius of 15$^\circ$ around NGC 4945. To check the effect of energy dispersion\footnote{http://fermi.gsfc.nasa.gov/ssc/data/analysis/documentation/ Pass8\_edisp\_usage.html}, which currently cannot be included in the unbinned analysis, we also performed the binned analysis, using $20^\circ \times 20^\circ$ square region centered on NGC 4945. We found that the dispersion correction insignificantly affects our results.

\begin{figure*}[t]
\centerline{\includegraphics[width=85mm]{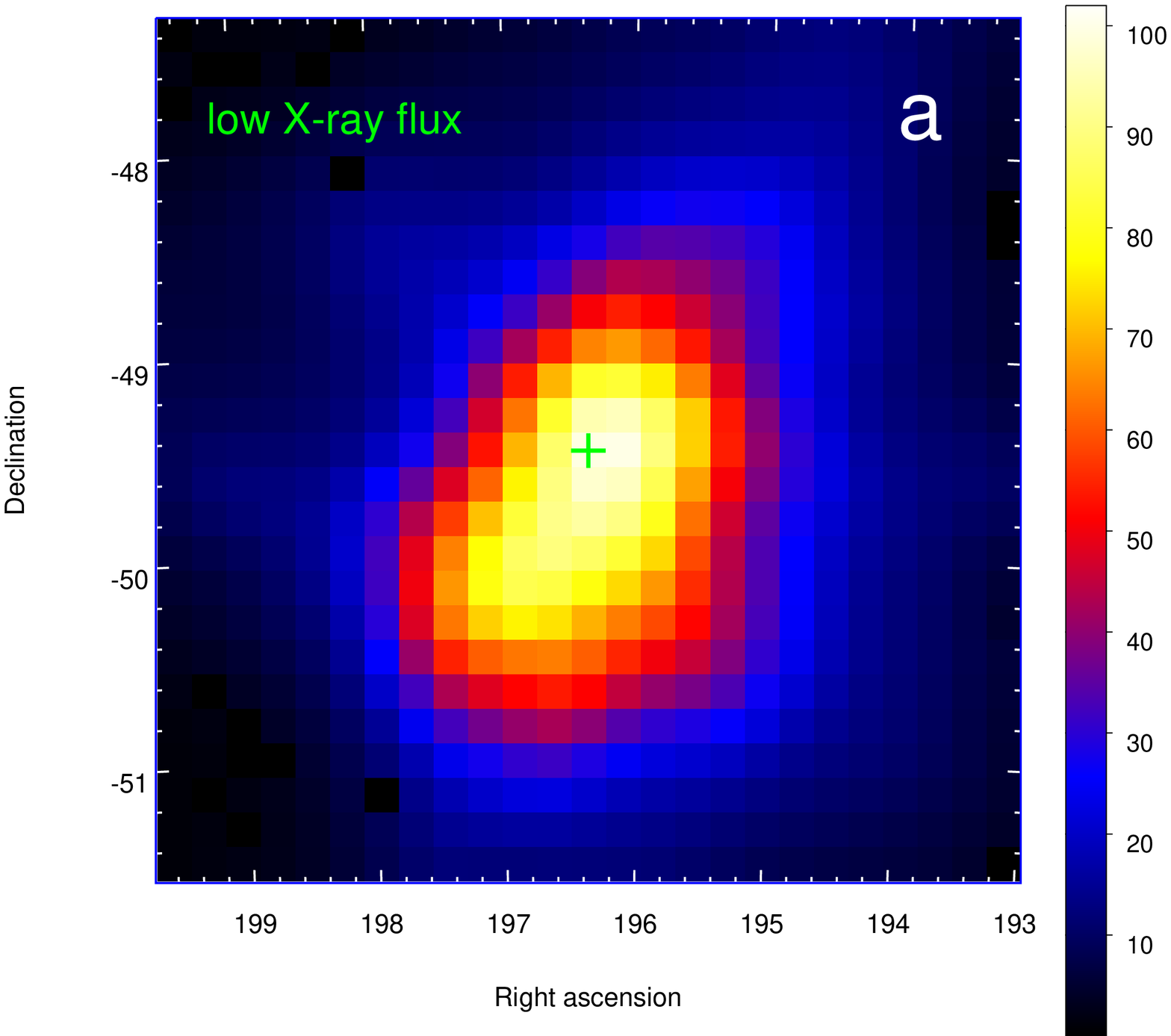} \includegraphics[width=85mm]{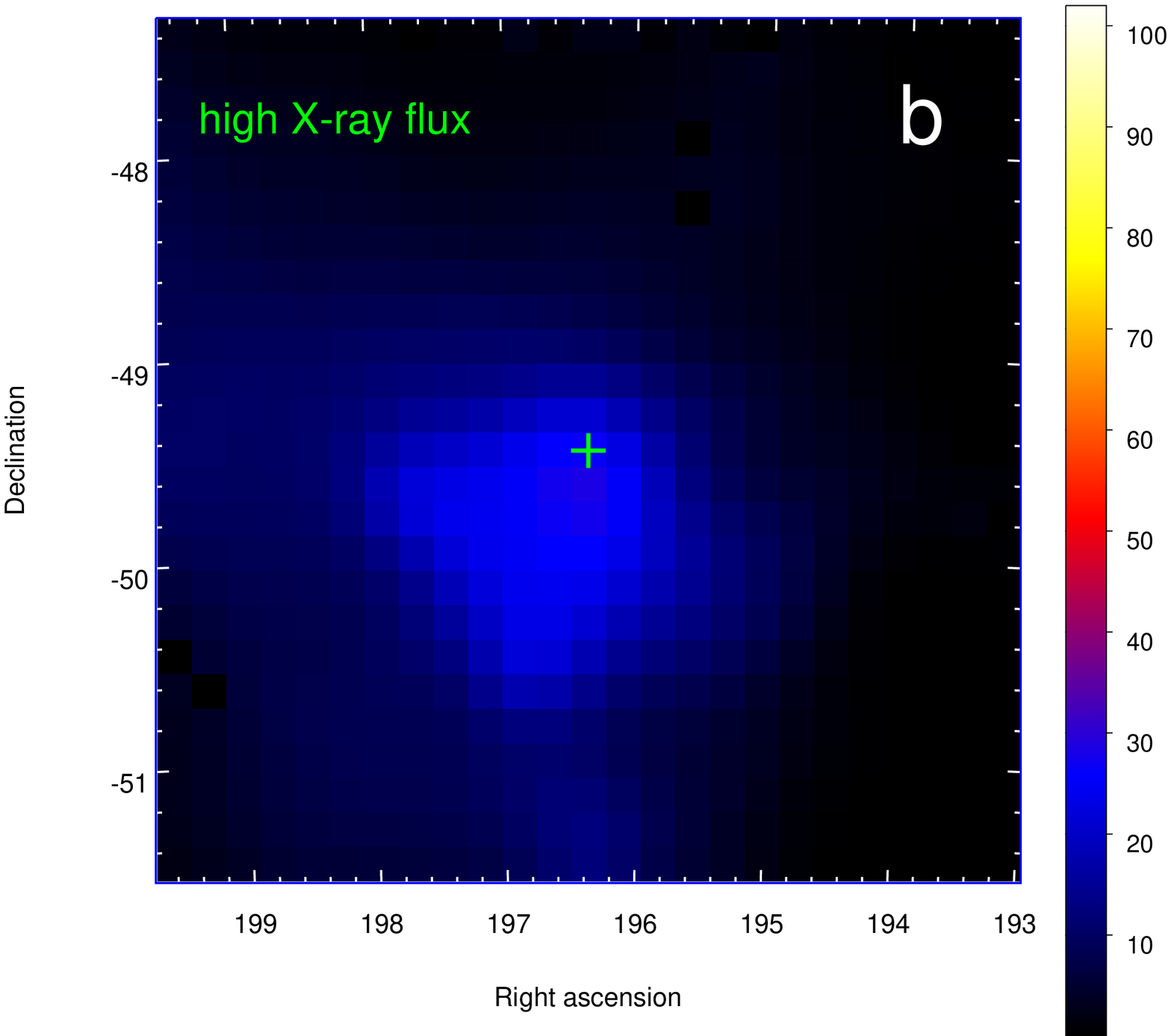}}
\centerline{\includegraphics[width=85mm]{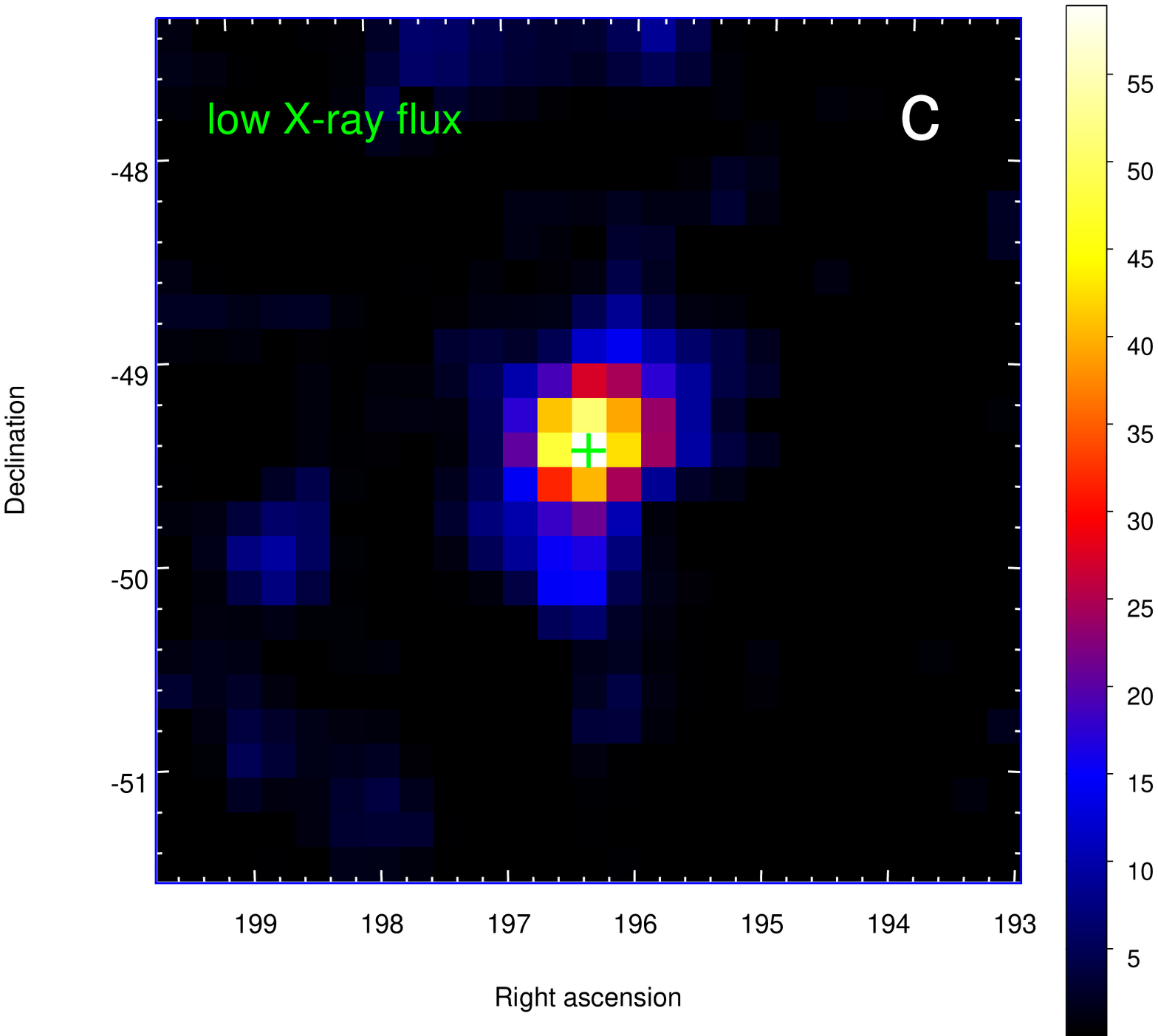} \includegraphics[width=85mm]{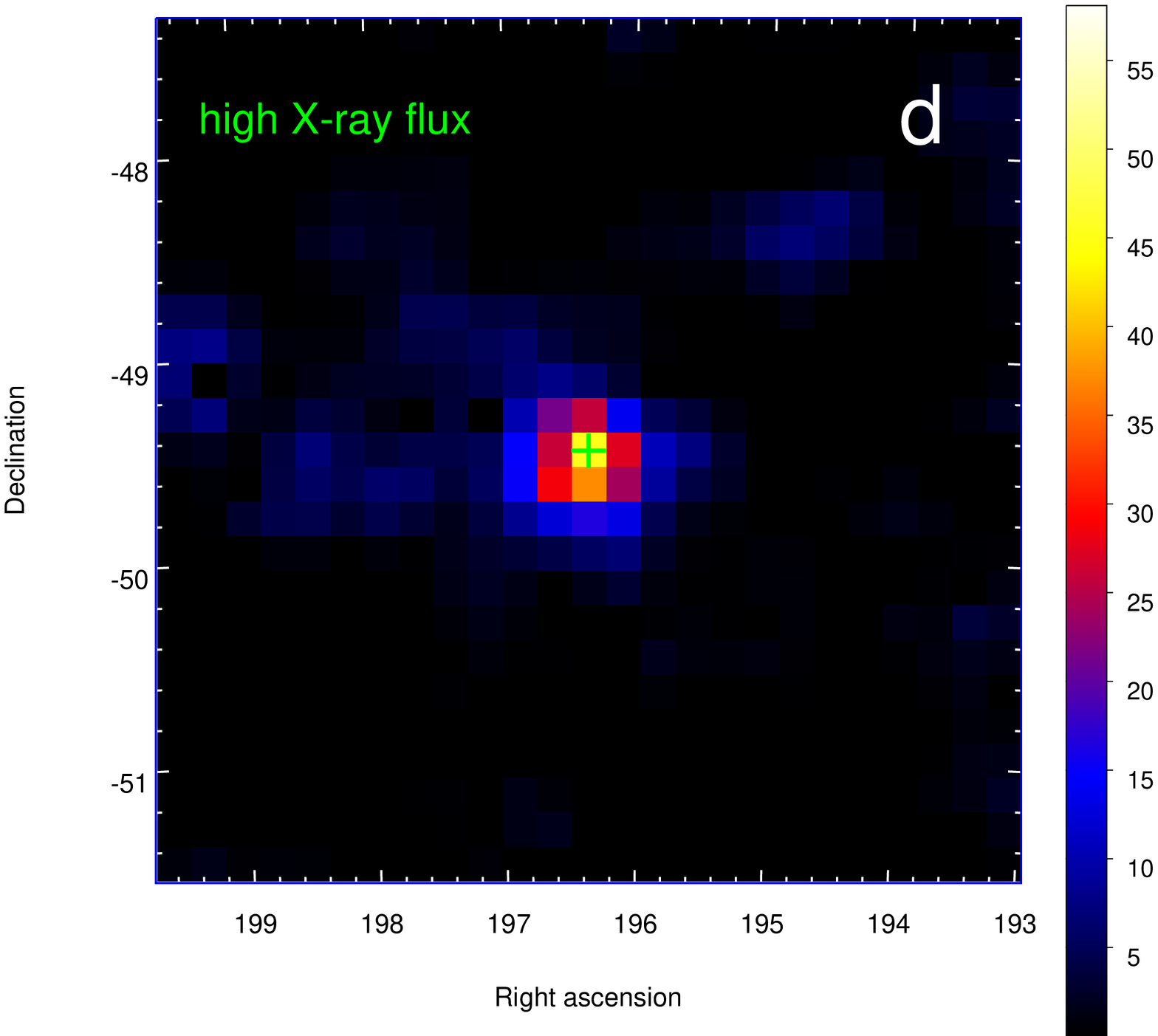}}
\centerline{\includegraphics[width=85mm]{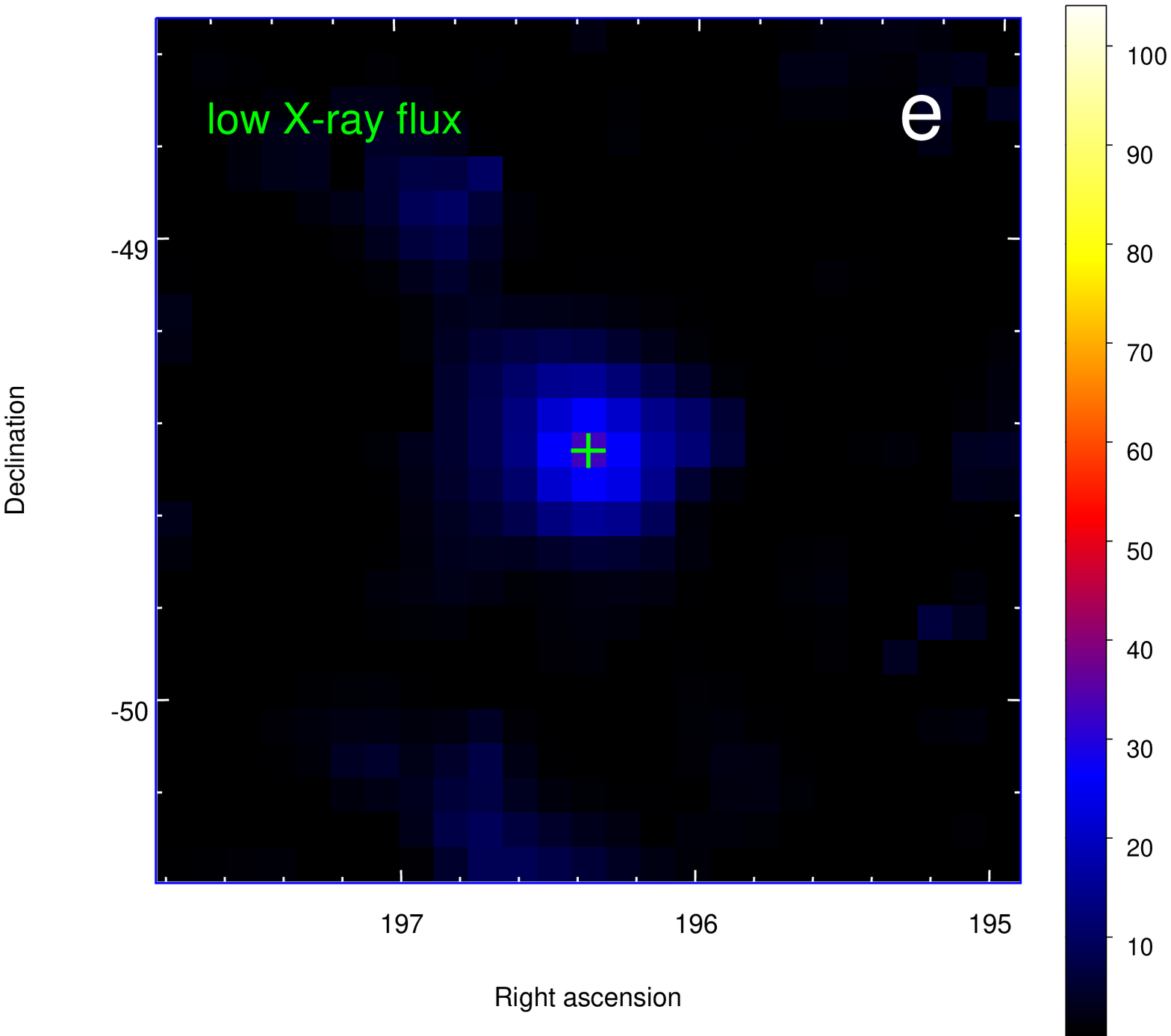} \includegraphics[width=85mm]{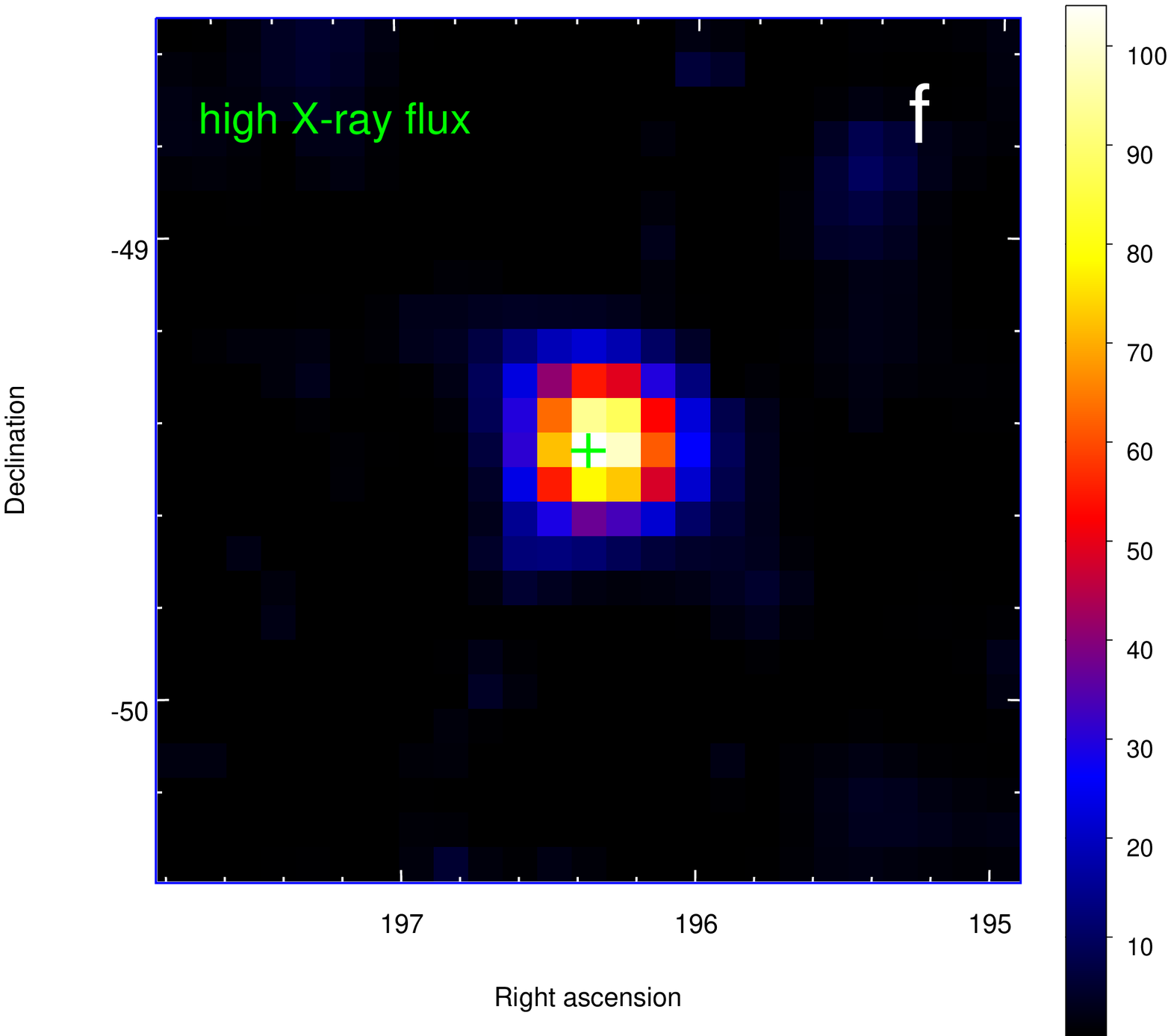}}
\caption{TS maps for the region around NGC 4945 built using the events detected during  the low (dataset L; left maps) and high (dataset H; right maps)  X-ray flux levels. Top maps are for events with $E<1$ GeV with a pixel size of $0\fdg17$; middle maps are for events with $E$ between 1 and 3 GeV with a pixel size of $0\fdg17$; bottom maps are for events with $E>3$ GeV with a pixel size of $0\fdg075$. In all panels the green cross shows the location of NGC 4945. In all panels, the 3FGL sources were  subtracted from the maps (except for 3FGL J1305.4-4926 which is the $\gamma$-ray counterpart of this galaxy).
} 
\label{fig:2}
\end{figure*}

In the TS maps of our region of interest, obtained by subtracting the best-fit model from the data, we do not find residuals indicating presence of unmodeled sources which could affect our analysis.  However, we note contamination of the $\gamma$-ray signal below 1 GeV, resulting in an apparent extension toward the bottom left corner in Figures \ref{fig:2}ab, by 3FGL J1251.0-4943 (the $\gamma$-ray counterpart of BL Lac object PMN J1326 5256) located at (R.A., Decl.) =  ($201\fdg7$, $-52\fdg9$). It is a factor of $\sim 2.6$ brighter in $\gamma$-rays than NGC 4945 and at the distance of $4\fdg8$ it may weakly contribute to the flux below 1 GeV measured in NGC 4945. The fitted parameters of this source are the same in all considered data sets, so its presence does not affect  our conclusions on spectral changes in NGC 4945.

\begin{figure}[t]
\centerline{\includegraphics[width=83mm]{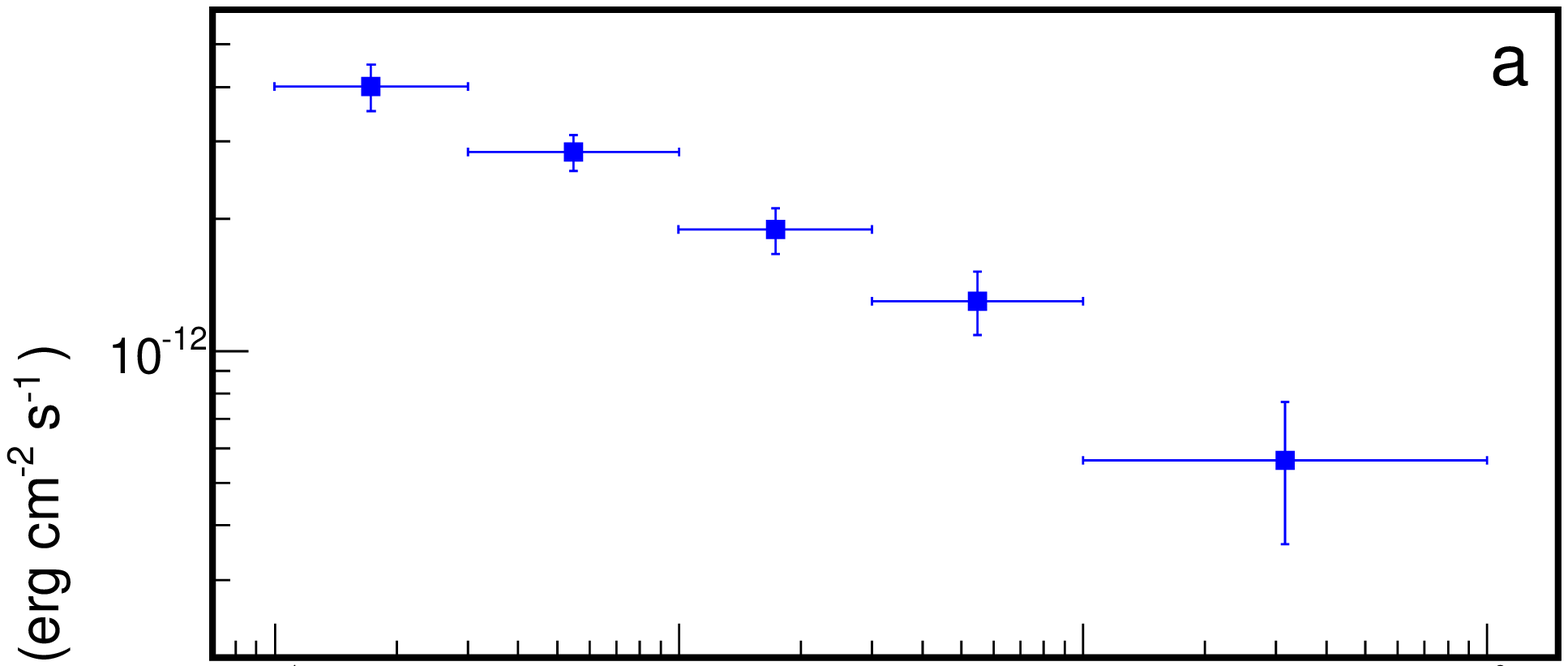}}
\centerline{\includegraphics[width=83mm]{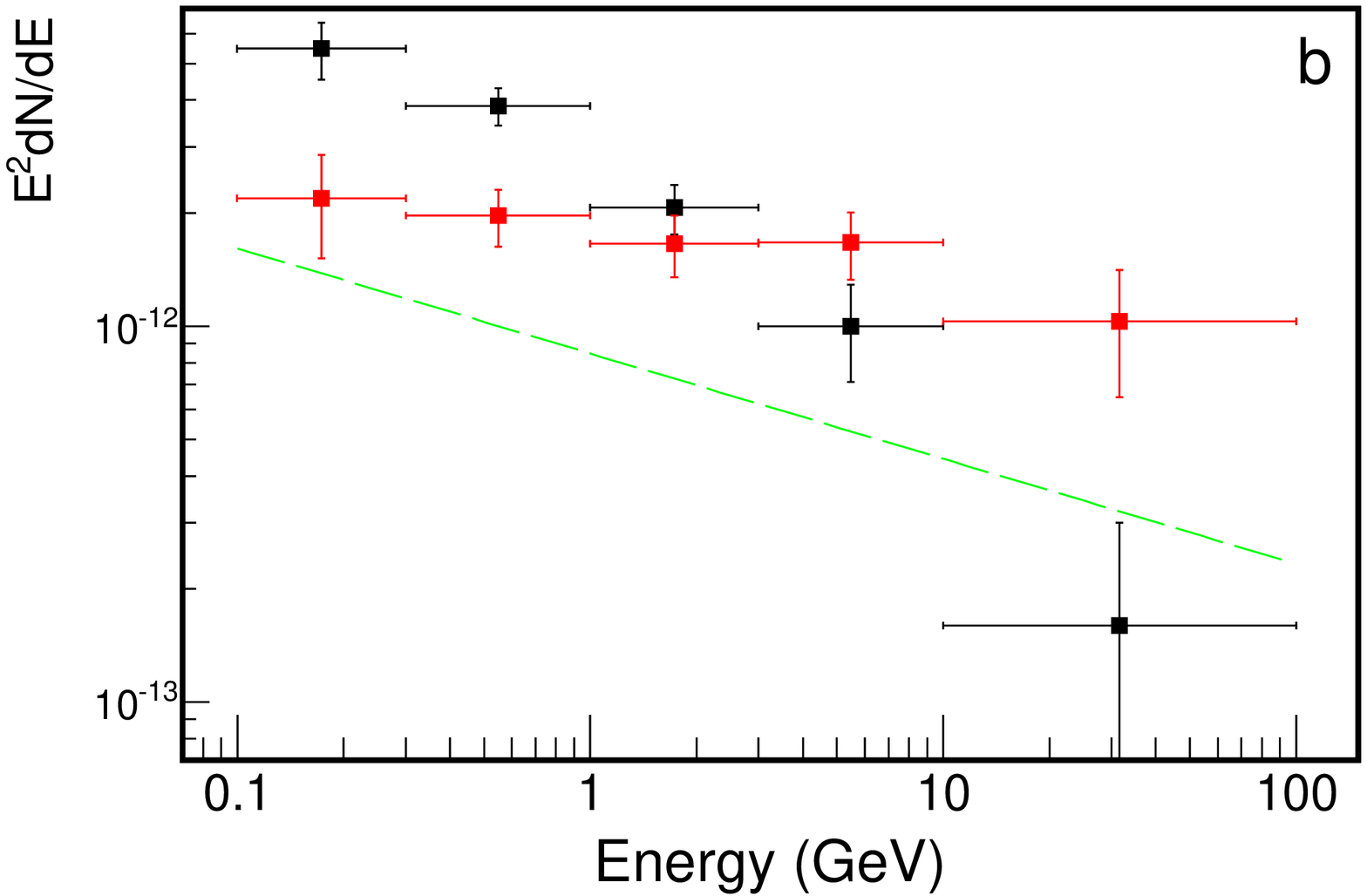}}
\caption{
NGC 4945 spectral energy distributions for the total LAT dataset T (panel a) and datasets resolved by the X-ray flux (panel b); SED for L is shown in black and SED for H is shown in red. The dashed green line in (b) represents the $\gamma$-ray spectrum of M 82 (the power-law fit in Table \ref{tab:PLfit}) scaled by the difference of IR luminosities (linear scaling with $L_{\rm IR}$ assumed) and distances between M 82 and NGC 4945.
} 
\label{fig:1}
\end{figure}

Results reported in Section \ref{sec:results} were obtained in models with spectral parameters of all sources adjusted to maximize the likelihood of the fit for a given dataset. The spectral energy distributions (SED) were obtained by fixing the spectral index, $\Gamma$, for NGC 4945 to the value from the power-law fit to a given dataset. However, to verify if variability in nearby sources could affect our results, we also repeated the analysis for datasets L, L5, L10, H, H5, and H10 using the model with spectral parameters of the background sources frozen to the values of the best-fit model to the total LAT data set (i.e.\ 2821 days; dataset T); below we refer to this variant as model T. For SEDs computed with model T we also fixed $\Gamma=2.33$ (from the power-law fit to dataset T) for NGC 4945. We found that it did not affect our results, i.e.\ using model T for the background sources we obtained parameters of the power-law fits consistent with those given in Table \ref{tab:PLfit} as well as TS maps consistent with those shown in Figure \ref{fig:2} (i.e.\ showing similar TS values at the position of NGC 4945). Also SEDs (Figure \ref{fig:1}b) are weakly affected by applying model T,  except for the lowest bin, 0.1--0.3 GeV, which shows a moderate dependence on the applied model. The difference of fluxes between L and H in this (0.1--0.3 GeV) bin decreases from a factor of $\simeq 2.5$ (in the best-fit models shown in Figure \ref{fig:1}) to $1.7$ (with model T).

\begin{table}[!t]
\begin{center}
\caption{Results of the power-law fits in the 0.1--100 GeV range to NGC 4945 using data sets defined in Section \ref{sec:data} (the number of days is given in parentheses) and to other starbursts and Seyfert 2 galaxies.
}
\label{tab:PLfit}
\begin{tabular}{|c|c|c|c|c|}
\hline
\cellspace 	& \cellspace 	$\overline{\mathcal F}_\mathrm{X}$ \cellspace 	& 	$\Gamma$ 	& 	$\mathcal F_\mathrm{\gamma}$ 	& 	TS  \\
\hline
 \multicolumn{5}{c}{ \cellspace NGC 4945 data sets  \cellspace  } \\
\hline
\cellspace L (1393) \cellspace   & 	$0.54 \pm 0.03$  		& 	$2.47 \pm 0.07$  &  	$2.6 \pm 0.3$ 				&  	221  \\
\hline
\cellspace H (1390) \cellspace 	& 	$2.87 \pm 0.04$			&	$2.11 \pm 0.08$  &  	$1.2 \pm 0.2$ 				&  	189    \\
\hline
\cellspace L5 (920) \cellspace 	& 	$0.57 \pm 0.04$  		& 	$2.46 \pm 0.09$  &  	$2.5 \pm 0.4$ 				&  	136  \\
\hline
\cellspace H5 (920) \cellspace  	& 	$2.75 \pm 0.04$			&	$2.05 \pm 0.09$  &  	$1.1 \pm 0.2$ 				&  	140    \\
\hline
\cellspace L10 (418) \cellspace 	& 	$0.60 \pm 0.04$  		& 	$2.56 \pm 0.10$  &  	$3.1 \pm 0.4$ 				&  	80  \\
\hline
\cellspace H10 (420) \cellspace  	& 	$2.66 \pm 0.05$			&	$2.07 \pm 0.11$  &  	$1.0 \pm 0.2$ 				&  	46    \\
\hline
\cellspace T (2821)  & 	$1.59 \pm 0.02$			&	$2.33 \pm 0.05$  &  	$1.9 \pm 0.2$ 				&  410 \\
\hline
 \multicolumn{5}{c}{ \cellspace  other objects  \cellspace}    \\
\hline
\cellspace M 82  & & \cellspace 	$2.28 \pm 0.04$  &  	$1.8 \pm 0.1$				&  	1010  \\
\hline
\cellspace NGC 253 \cellspace 	& &	$2.14 \pm 0.05$  &  	$1.1 \pm 0.1$			&  	616    \\
\hline
\cellspace Circinus \cellspace 
	& &	$2.43 \pm 0.09$  &  	$2.0 \pm 0.3$ 	&  	103  \\
\hline
\cellspace NGC 1068 \cellspace 
& &	$2.47 \pm 0.05$  &  	$1.6 \pm 0.1$ 	&  357    \\
\hline
\end{tabular}
\end{center}
$\overline{\mathcal F}_\mathrm{X}$ is the average BAT count rate in the 15--50 keV range in units of $10^{-3}$ cts cm$^{-2}$ s$^{-1}$; $\Gamma$ is the $\gamma$-ray  power-law photon index and $\mathcal F_\mathrm{\gamma}$ is the photon flux in the 0.1--100 GeV range
in units of  $10^{-8}$ ph cm$^{-2}$ s$^{-1}$.
\end{table}

For a comparison of $\gamma$-ray loud galaxies in Section \ref{sect:disc},  we analyzed the LAT data from 8 years of observations of NGC 253, M 82, NGC 1068 and Circinus; in the model for NGC 1068, in addition to 3FGL sources, we included the new $\gamma$-ray source, at the distance of $\sim 4^\circ$, found in \citet{2016A&A...596A..68L}. Our luminosities in the 0.1--100 GeV range,   $L_{\gamma}$, for NGC 4945 and NGC 1068, placing them at/above the calorimetric limit in Figure \ref{fig:6}, are by a factor of 1.5 and 1.7, respectively, larger than found in \citet{2012ApJ...755..164A}. The difference results from analysis improvements in Pass 8 data with respect to Pass 7 data used by \citet{2012ApJ...755..164A}. For NGC 1068 we note a marginal hint for the change of the $\gamma$-ray spectrum. For the first 3 years of LAT observations, the 
same as used in  \citet{2012ApJ...755..164A}, we found $\Gamma=2.37 \pm 0.09$ and $\mathcal F_\mathrm{\gamma} = (1.3 \pm 0.2) \times 10^{-8}$ ph cm$^{-2}$ s$^{-1}$, whereas for the last four years of our dataset (i.e.\ after August 2012), $\Gamma= 2.51 \pm 0.08$ and $\mathcal F_\mathrm{\gamma} = (1.7 \pm 0.2) \times 10^{-8}$ ph cm$^{-2}$ s$^{-1}$.

\section{Results}
\label{sec:results}

Our main results are presented in Figures \ref{fig:2} and \ref{fig:1} and Table \ref{tab:PLfit}. All our results indicate that the $\gamma$-ray spectrum of NGC 4945 changes with the change of its hard X-ray flux. The significance of the spectral difference between the power-law fits for datasets H and L is $\simeq 5 \sigma$. SEDs shown in Figure \ref{fig:1}b indicate that a large difference of the detection significance between H and L should be seen both below 1 GeV and above 3 GeV. Indeed, with an equal exposure time in both datasets, the change of the $\gamma$-ray signal is seen in the TS maps (see Figure \ref{fig:2}) for both 0.1--1 GeV and 3--100 GeV with a formal statistical significance of $\simeq 9 \sigma$ (determined as $\sqrt{| {\rm TS}_{\rm L} - {\rm TS}_{\rm H} |}$, where ${\rm TS}_{\rm X}$ is the test statistic at the position of NGC 4945 in dataset X). Note that these results, indicating the change of the $\gamma$-ray signal at low and high energies, are mutually  independent. Our results for data sets L5/H5 and L10/H10 are consistent with those shown in Figures \ref{fig:2} and \ref{fig:1}.

\begin{figure}[t]
\centerline{\includegraphics[width=87mm]{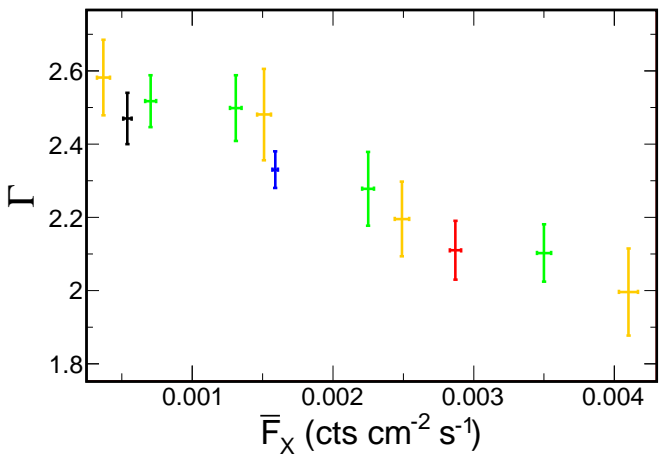}}
\caption{Photon spectral index of power-law fits in the 0.1--100 GeV range as a function of $\overline{\mathcal F}_\mathrm{X}$. The yellow points are for datasets containing 620 days in consecutive, non-overlapping ranges of $\mathcal F_\mathrm{X}$. The green points are for datasets containing 900 days in similar but partially overlapping ranges of $\mathcal F_\mathrm{X}$. The black, blue and red point is for dataset L, T and H.
} 
\label{fig:3}
\end{figure}

We also performed the division of LAT observations into datasets corresponding to smaller ranges of $\mathcal F_\mathrm{X}$ than those of L and H.  Again, we find a systematic indication of hardening of the $\gamma$-ray spectrum with the increase of the X-ray flux, see Figure \ref{fig:3}, although obviously the statistical uncertainty on spectral parameters increases with a decreasing size of datasets. Using all fitting results shown in Figure \ref{fig:3} we find the Pearson correlation coefficient for the $\Gamma$--$\overline{\mathcal F}_\mathrm{X}$ relation, weighted by the inverse of spectral index uncertainty, of $r \simeq -0.97$ with the $p$-value of $\simeq 2 \times 10^{-6}$.

\begin{figure}[t]
\centerline{\includegraphics[width=85mm]{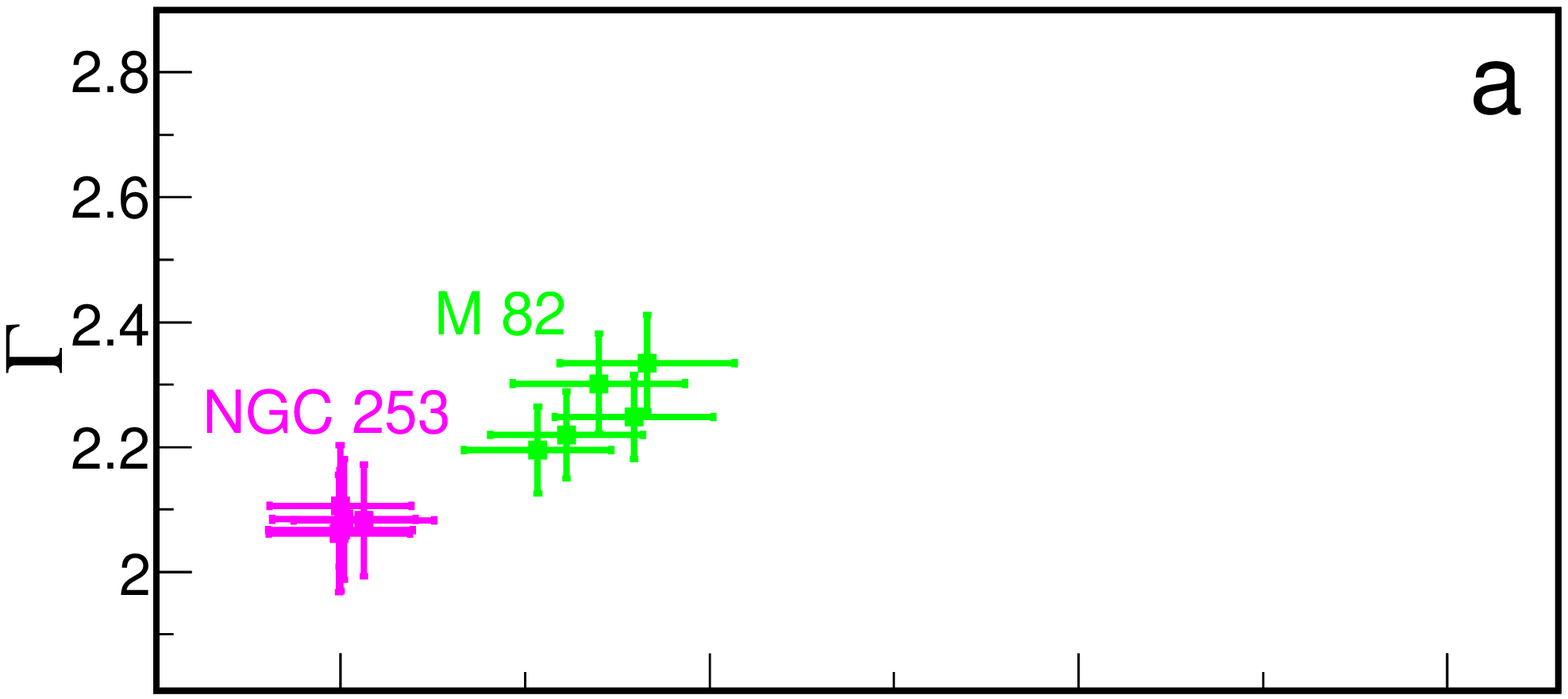}}
\centerline{\includegraphics[width=85mm]{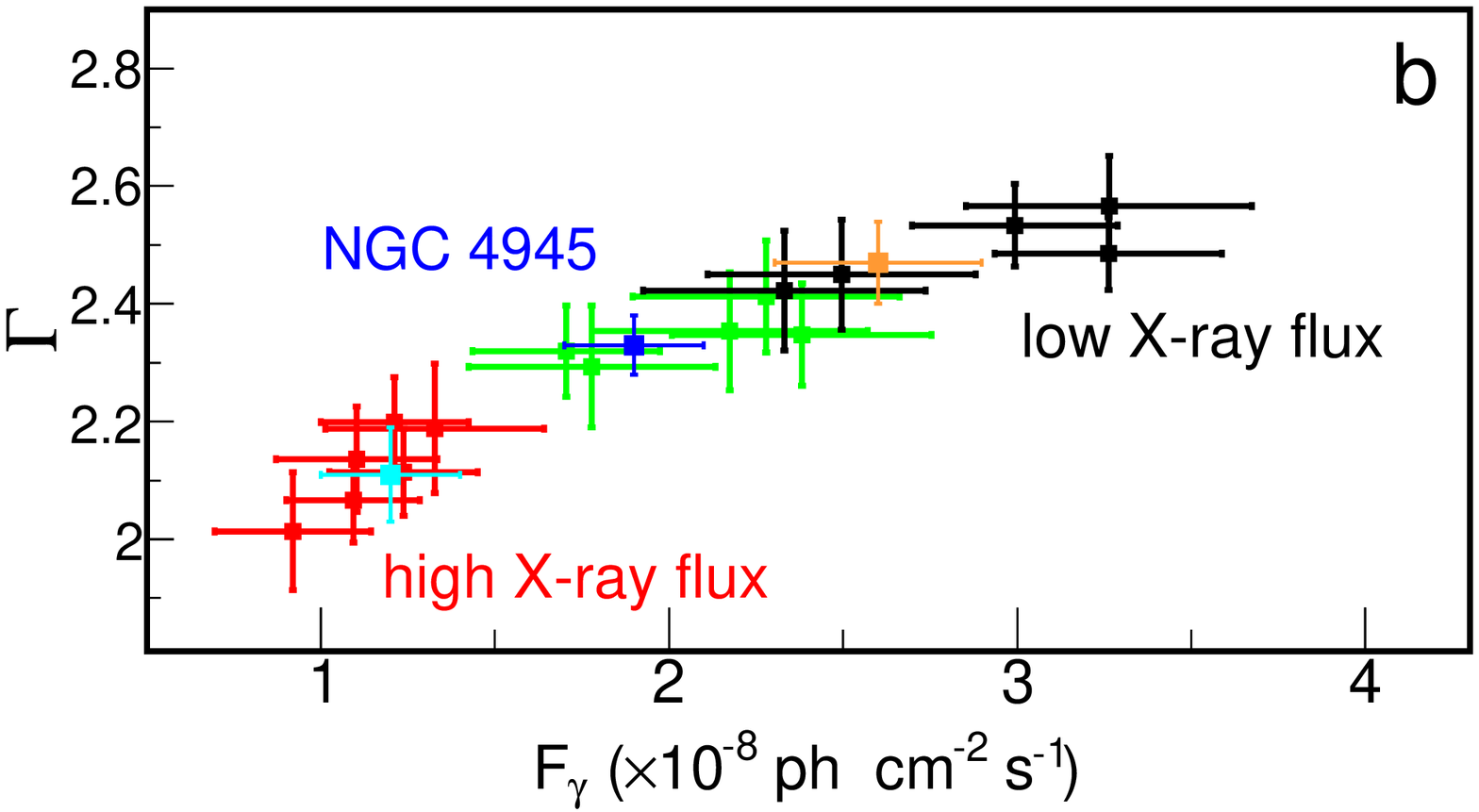}}
\caption{Parameters of the power-law fits in the 0.1--100 GeV range. Panel (a) is for datasets comprising 800 days randomly drawn from the total LAT dataset of NGC 253 (magenta) and M 82 (green). Panel (b) is for datasets comprising 800 days randomly drawn from set T (green), L (black) and H (red) in NGC 4945. The blue, orange and cyan points in (b) show parameters of power-law fits to dataset T, L and H, respectively, as given in Table \ref{tab:PLfit}.
} 
\label{fig:4}
\end{figure}

We have made a number of tests for LAT datasets selected without the X-ray flux criterion and in all cases we found that they are consistent, within uncertainties, with the results for the total LAT dataset T. An example is shown in Figure \ref{fig:4}b, were power-law fits to datasets of 800 days randomly drawn from set T are compared with fits to datasets of 800 days randomly drawn from set L or H. We see that the latter, i.e.\ black and red points, occupy two separate areas in the parameter plane. The green points for datasets mixing various X-ray flux levels have a larger spread of fitted parameters than those for NGC 253 and M 82 (whose $\gamma$-ray signals should be constant in time; Figure \ref{fig:4}a), reflecting larger uncertainties on $\Gamma$ in 800-day datasets for NGC 4945. However, it is much smaller than the overall spread of parameters for $\mathcal F_\mathrm{X}$-resolved datasets.

Using additional 10 datasets of 400 days drawn from set L and 10 such datasets drawn from H, which again give parameters distributed in  separate areas in the parameter plane similar to that shown in Figure \ref{fig:4}b, we find that these two samples (i.e.\ drawn from either L or H) are significantly different, with the {\it p}-value of $\simeq 10^{-6}$ from the two-dimensional Kolmogorov-Smirnov test \citep{1987MNRAS.225..155F,1992nrfa.book.....P}. Although dataset L or H could be affected by a background fluctuation or short time-scale outbursts of nearby sources (on a time scale of days, longer outbursts should give similar contribution to both L and H), it is highly unlikely that such a fluctuation/outburst affects also all these shorter data sets in a manner illustrated in Figure \ref{fig:4}.

\section{Discussion}
\label{sect:disc}

Apart from  NGC 4945, two other Seyfert 2 galaxies, NGC 1068 and Circinus, have been detected by LAT \citep{2013ApJ...779..131H,2010A&A...524A..72L}. Similar to NGC 4945, these galaxies exhibit a composite starburst/AGN activity and the interpretation of their $\gamma$-ray emission is uncertain. Among the three $\gamma$-ray loud Seyfert 2s, the X/$\gamma$-ray correlation can be investigated only in NGC 4945, which  has the largest $\gamma$-ray detection significance, and whose variable X-ray emission  from the nucleus can be directly observed. In contrast, the X-ray radiation from NGC 1068 is fully reflection dominated and no variability is observed. Circinus, in turn, is strongly contaminated by the Galactic plane and its detection significance is too low to search for changes of the $\gamma$-ray spectrum.

The correlation revealed in our study implies that the $\gamma$-ray production in NGC 4945 is dominated by the active nucleus. On the other hand, at the supernova rate estimated for NGC 4945 \citep{2009AJ....137..537L}, the injected  cosmic ray power is sufficient to produce the observed $L_{\gamma}$ through $\pi^0$-decay emission \citep[cf.][]{2010A&A...524A..72L,2016ApJ...821...87E}. Below we briefly discuss this issue in the context of $\gamma$-ray observations of other Seyfert and starburst galaxies. 

The GeV and TeV detections of two nearby starburst galaxies, M82 and NGC 253 \citep{2009Sci...326.1080A,2009Natur.462..770V,2010ApJ...709L.152A}, have confirmed the link between the star-formation activity and the $\gamma$-ray emission likely related to pionic interactions of cosmic rays  with the interstellar medium; a universal scaling of $L_\gamma$ with the star-formation rate  was then proposed in \citet{2012ApJ...755..164A}. However, an efficient  $\gamma$-ray emission in this class of objects appears to be not as ubiquitous as could be expected after these first detections  \citep[cf.][]{2016MNRAS.463.1068R} indicating that non-radiative losses (i.e.\ diffusive or advective escape) dominate in some starbursts.

\begin{figure}[t]
\centerline{\includegraphics[width=87mm]{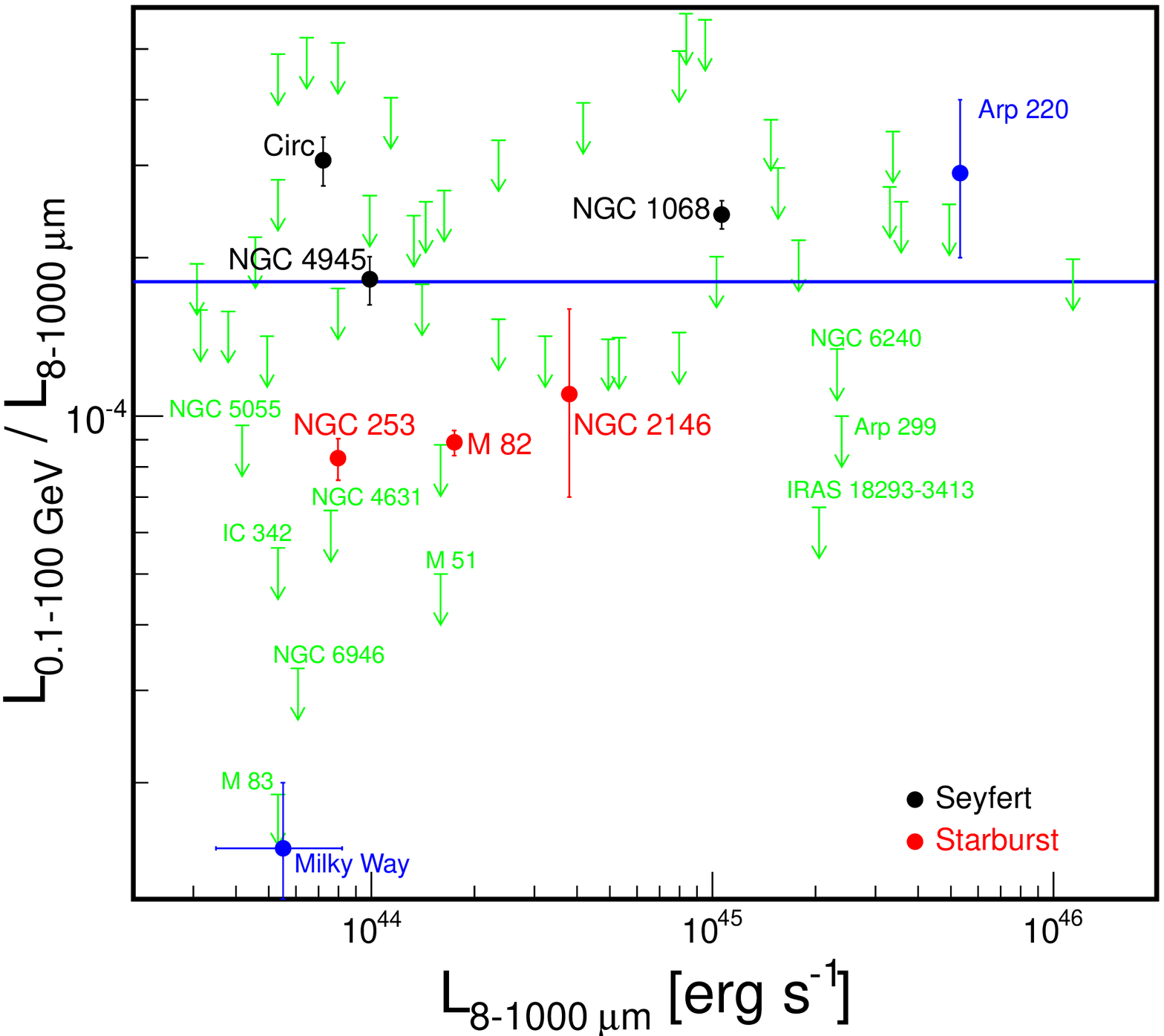}}
\caption{Comparison of the $\gamma$-ray (0.1--100 GeV) and IR (8--1000 $\mu$m) luminosities for star-forming and Seyfert galaxies. We adopted data from \citet{2014ApJ...794...26T} for NGC 2146,  \citet{2016ApJ...823L..17G} for Arp 220, \citet{2012ApJ...755..164A} for IR luminosities and Milky Way $\gamma$-ray luminosity, upper limits (green arrows) from \citet{2016MNRAS.463.1068R} for non-detected galaxies and we used our results for NGC 4945 (dataset T), NGC 1068, Circ, M 82 and NGC 253 (Table \ref{tab:PLfit}). The blue line shows the calorimetric limit.
} 
\label{fig:6}
\end{figure}

Figure \ref{fig:6} compares the $\gamma$-ray luminosity of star-forming galaxies with their IR luminosity, which is a good tracer of the star-formation rate \citep[e.g.][]{1998ApJ...498..541K} and, hence, estimates the injected cosmic ray power. The blue line corresponds to the calorimetric limit, in which the total energy of cosmic rays is used for the pion production and assuming that each supernova injects $10^{50}$ ergs of cosmic ray energy \citep[e.g.,][]{2012ApJ...755..164A}. The $\gamma$-ray luminosities of starburst galaxies, M 82 and NGC 253, a factor of $\sim 2$ below this limit, indicate that $\sim 50$\% of cosmic ray energy is channeled for pion production, as also found in the starburst models for these objects \citep[e.g.,][]{2011ApJ...734..107L,2014ApJ...780..137Y}. These starbursts have harder $\gamma$-ray spectra than the Seyfert galaxies, so the (energy-dependent) diffusive losses would be more important in the starburst scenario for the latter. Yet, Seyferts are at or above the calorimetric limit, which presents a major problem for the starburst model \citep[cf.][]{2013ApJ...779..131H} and points toward a significant contribution from their  AGN components. Studies of $\gamma$-ray  emission from NGC 1068 \citep[e.g.,][]{2010A&A...524A..72L,2014ApJ...780..137Y,2016ApJ...821...87E} indeed find that it cannot be explained by the starburst activity  and indicate the active nucleus as the primary source of $\gamma$-rays. 

A constant in time contribution of the starburst component in NGC 4945, which obviously cannot exceed the level of $\gamma$-ray emission found in X-ray resolved data sets, can still significantly contribute to the total $L_{\gamma}$. For parameters similar to those of M 82 (the dashed line in Figure \ref{fig:1}b) such a diffuse emission would only slightly overpredict the flux above 10 GeV in dataset L. Then, with a small reduction of the $\gamma$-ray production efficiency, it could fully account for emission above 10 GeV measured for low X-ray flux levels and give a $\sim 20$\% contribution to the total $L_{\gamma}$. For much lower efficiencies of $\gamma$-ray production, e.g.\ such as assessed for M 83, which has a high surface density \citep[but lower than NGC 4945; cf.][]{2011ApJ...734..107L}, contribution from the sturburst component would be insignificant.

The Seyfert galaxies exhibit also an interesting similarity of their Eddington-scaled X-ray and $\gamma$-ray luminosities \citep[see figure 1 in][]{2015A&A...584A..20W}, with $L_{\gamma}/L_{\rm Edd} \simeq 10^{-4}$ (0.1--100 GeV) and $\lambda_{\rm X} \equiv L_{\rm X}/L_{\rm Edd} \simeq 0.02$ (2--10 keV) in all three objects. These are the largest values of $\lambda_{\rm X}$ observed in nearby AGNs. At such $\lambda_{\rm X}$ transition between the hard and soft spectral state is observed in black-hole binaries; softening of the X-ray spectrum with increasing luminosity observed in  NGC 4945 \citep{2014ApJ...793...26P,2012A&A...537A..87C} resembles the behavior of black-hole transients during this spectral state change.

A prominent $\gamma$-ray signal has not been detected from other radio-quiet Seyfert galaxies \citep{2012ApJ...747..104A}. However, in the context of high-energy emission from accretion flows, the available LAT upper limits provide only a  moderate  constraint of $\la 10$\% on the fraction of accretion power which can be used for acceleration of protons and/or electrons emitting in the GeV range \citep{2015A&A...584A..20W}. Presence of such nonthermal particles in accretion flows can be expected both from a theoretical point of view \citep[e.g.,][]{2016ApJ...822...88K} and from modeling of some observed spectra \citep[e.g.,][]{2014SSRv..183...61P}, but any $\gamma$-ray radiation produced within accretion flow, or in its vicinity, would be strongly  attenuated by $\gamma \gamma$ interactions with the ambient radiation field.

We used the model of internal $\gamma \gamma$ absorption, developed in our previous works \citep{2013MNRAS.432.1576N,2015A&A...584A..20W}, to assess possible locations of the $\gamma$-ray  emitting site in the NGC 4945 active nucleus. Geometry of the inner accretion flow and the nature of the X-ray source are highly uncertain for NGC 4945 due to the obscuration of its nucleus. We considered the case of an optically thick disk extending down close to the black hole, motivated by the commonly accepted model of black-hole binaries  relating transition to the soft spectral state, at $L \sim 0.1 L_{\rm Edd}$, with rebuilding of such a disk \citep[e.g.,][]{2007A&ARv..15....1D,2016AN....337..391M}.

Due to paucity of photons with $E > 10$ GeV we were not able to rule out or confirm a presence of a spectral break at these energies.  Using {\tt gtsrcprob} we found that the largest photon energy detected from NGC 4945 with high probability  ($> 80$\%) is $E \simeq 40$ GeV. Below we assume that the $\gamma$-ray spectrum of the high X-ray flux levels does not possess a high-energy cut-off up to at least a few tens of GeV. This requires the source to be at least $\sim (10^3 - 10^4) R_{\rm g}$ away from the inner accretion disk, depending on inclination, where $R_{\rm g}=GM/c^2$ is the gravitational radius; the distance may be lower than $10^3 R_{\rm g}$ if we observe the accretion disk from a face-on direction, which is unlikely for a Seyfert 2 galaxy. We assumed that the $\gamma$-ray source is located at the symmetry axis, and we used the standard disk model of \cite{1973A&A....24..337S}. To assess the maximum effect of $\gamma$-ray attenuation in thermal-radiation field of such a disk, we assumed that it accretes with the rate $\dot M = L_{\rm Edd}/c^2$, and extends down to the innermost stable circular orbit (ISCO) at $6 R_{\rm g}$. We took into account a compact, centrally located X-ray source matching the internal X-ray luminosity and spectral index reported in \cite{2014ApJ...793...26P}; the exact geometry of this source is not important for opacity to $\gamma$-rays emitted far from it.
Due to anisotropy of the radiation field, the cut-off energy related with $\gamma \gamma$ absorption depends on the viewing angle, see e.g.\ \citet{2011A&A...529A.120C}, but note that the inner disk of the binary system considered in that work emits in soft X-rays whereas that in NGC 4945 nucleus would emit in UV, so quantitative results are different. 

On the other hand, by the causality argument, the $\gamma$-ray source located more than $\sim 1$ light day away from the X-ray source would not be able to respond  to its changes in the manner indicated by our results. Again assuming that the X-rays are produced by a compact source close to the black hole, this constrains the distance of the $\gamma$-ray source to  $\la 10^4 R_{\rm g}$. This, in turn, rules out edge-on observing directions, with inclination angles $\ga 70\degr$, for which a larger distance is required by the $\gamma$-ray transparency condition. We note also that a detection of NGC 4945 above 100 GeV, with a power-law spectrum extending from the LAT range, would contradict our findings, as photons with such energies cannot escape from the region within 1 light day. In this context it is interesting to note that NGC 4945 (as well as the other two Seyferts) has not been detected in this range so far, although a detection is within reach of currently operating imaging atmospheric Cherenkov telescopes. 

A $\gamma$-ray source located $\sim (10^3 - 10^4) R_{\rm g}$ from the central engine would be possibly related with formation of a weak jet. A jet-like structure is indeed observed in NGC 4945 nucleus, and also in NGC 1068 and Circinus, but such nuclear jets are commonly found in other Seyfert galaxies  as well \citep[e.g.][]{2006AJ....132..546G,1998MNRAS.297.1202E,2009AJ....137..537L}. The specific conditions underlying the $\gamma$-ray loudness of NGC 4945, NGC 1068 and Circinus may then involve their high $L/L_{\rm Edd}$ values and the related changes in accretion flow. In particular, it may involve a rapid increase of the jet velocity occurring when an inwards-moving inner disk edge in a source making a hard to soft state transition  approaches  the ISCO, which is the preferred interpretation of the disc-jet coupling, well-established in black-hole X-ray binary systems \citep[e.g.][]{2004MNRAS.355.1105F} and observed also in AGNs \citep[e.g.][]{2002Natur.417..625M}. Particles would then be accelerated by an internal shock formed in the collision of a fast jet with a previously existing slower outflow.
Notably, \cite{2011A&A...529A.120C} estimate a similar distance (in units of $R_{\rm g}$) of the  $\gamma$-ray source for Cyg X-3, where the $\gamma$-ray emission is related with hard/soft state transitions \citep[e.g.][]{2012MNRAS.421.2947C} occurring at $L/L_{\rm Edd}$ similar to that of NGC 4945.

The $\gamma$-ray spectrum of  the low X-ray flux levels can be observed from a source located at a smaller distance, $\sim 10^2 R_{\rm g}$; more precise estimates depend on the assumed geometry. E.g., the $\gamma$-rays may come from an inner optically-thin flow, located within a few tens of $R_{\rm g}$, if the starburst activity of this galaxy accounts for the emission above $\sim 10$ GeV (see above). We note that similar signals,  pronounced only below $\sim 1$ GeV, cannot be excluded in other Seyfert galaxies, as presence of nearby sources often does not allow for a proper assessment of a signal at such energies, as e.g.\ in the case of the brightest in hard X-rays Seyfert galaxy NGC 4151 \citep{2015A&A...584A..20W}.

\section{Summary}

We presented a novel approach to investigate the $\gamma$-ray variability in NGC 4945 by analyzing the LAT data selected based on the X-ray flux level. The $\gamma$-ray spectrum appears to be correlated with the X-ray luminosity, with changes of the $\gamma$-ray signal independently seen at low and high $\gamma$-ray energies. The X/$\gamma$-ray correlation is indicated by all datasets (comprising between $\sim 1$ and 4 years of LAT data) selected using the X-ray flux criterion, while datasets neglecting this criterion  are consistent with representing a non-varying $\gamma$-ray emission. We have thoroughly tested the dependence of our results on the approach to data analysis. 

The correlation implies that dominating contribution to the observed $\gamma$-ray emission comes from the active nucleus of NGC 4945 and this constrains the efficiency of $\gamma$-ray production related with starburst activity. The implied limit on the radiative efficiency (with $\la 20$\% of the cosmic ray power lost in pionic interactions, if the IR luminosity is used as a measure of the star-formation rate) is slightly lower than the efficiencies assessed for NGC 253 and M 82.

The nature of the $\gamma$-ray source may be different at low and high X-ray luminosities. At the latter, the $\gamma$-ray transparency and the causality conditions require the source to be located $\sim (10^3 - 10^4) R_{\rm g}$ away from the central black hole, if an inner optically-thick disk is present and the X-ray source is close to the black hole. We speculate that such a $\gamma$-ray emitting site may appear as a result of an inwards collapse of accretion disk, associated with the increase of luminosity. Then, it may manifest the disk-jet connection established in other accreting systems. At low X-ray luminosities,  the source may be located much closer to the black hole.

We noted similarities between NGC 4945, NGC 1068 and Circinus (similar Eddington ratios of high-energy emission, lack of TeV detections, unlikely high  efficiencies of $\gamma$-ray production in starburst scenario) which we regard as a further argument for a dominating contribution of their active nuclei to the $\gamma$-ray emission.

\acknowledgments

We thank the referee for valuable suggestions and Francesco Longo, Rachele Desiante and Jean Ballet for help with the Fermi/LAT data analysis. We made use of data and software provided by the Fermi Science Support Center, managed by the HEASARC at the Goddard Space Flight Center, and Swift/BAT transient monitor results provided by the Swift/BAT team. This work has been supported by the Polish National Science Center ETIUDA doctoral grant DEC-2016/20/T/ST9/00386 and 
OPUS grant DEC-2016/21/B/ST9/02388. \newline

\bibliographystyle{apj}

\begin{thebibliography}{}
\expandafter\ifx\csname natexlab\endcsname\relax\def\natexlab#1{#1}\fi
\providecommand{\url}[1]{\href{#1}{#1}}

\bibitem[{{Abdo} {et~al.}(2010{\natexlab{a}}){Abdo}, {Ackermann}, {Ajello},
  {Allafort}, {Antolini}, {Atwood}, {Axelsson}, {Baldini}, {Ballet},
  {Barbiellini}, \& et~al.}]{2010ApJS..188..405A}
{Abdo}, A.~A., {Ackermann}, M., {Ajello}, M., {et~al.} 2010{\natexlab{a}},
  \apjs, 188, 405

\bibitem[{{Abdo} {et~al.}(2010{\natexlab{b}}){Abdo}, {Ackermann}, {Ajello},
  {Atwood}, {Axelsson}, {Baldini}, {Ballet}, {Barbiellini}, {Bastieri},
  {Bechtol}, {Bellazzini}, {Berenji}, {Bloom}, {Bonamente}, {Borgland},
  {Bregeon}, {Brez}, {Brigida}, {Bruel}, {Burnett}, {Caliandro}, {Cameron},
  {Caraveo}, {Casandjian}, {Cavazzuti}, {Cecchi}, {{\c C}elik}, {Charles},
  {Chekhtman}, {Cheung}, {Chiang}, {Ciprini}, {Claus}, {Cohen-Tanugi},
  {Conrad}, {Dermer}, {de Angelis}, {de Palma}, {Digel}, {Silva}, {Drell},
  {Drlica-Wagner}, {Dubois}, {Dumora}, {Farnier}, {Favuzzi}, {Fegan}, {Focke},
  {Foschini}, {Frailis}, {Fukazawa}, {Funk}, {Fusco}, {Gargano}, {Gasparrini},
  {Gehrels}, {Germani}, {Giebels}, {Giglietto}, {Giordano}, {Glanzman},
  {Godfrey}, {Grenier}, {Grondin}, {Grove}, {Guillemot}, {Guiriec}, {Hanabata},
  {Harding}, {Hayashida}, {Hays}, {Hughes}, {J{\'o}hannesson}, {Johnson},
  {Johnson}, {Johnson}, {Kamae}, {Katagiri}, {Kataoka}, {Kawai}, {Kerr},
  {Kn{\"o}dlseder}, {Kocian}, {Kuss}, {Lande}, {Latronico}, {Lemoine-Goumard},
  {Longo}, {Loparco}, {Lott}, {Lovellette}, {Lubrano}, {Madejski}, {Makeev},
  {Mazziotta}, {McConville}, {McEnery}, {Meurer}, {Michelson}, {Mitthumsiri},
  {Mizuno}, {Moiseev}, {Monte}, {Monzani}, {Morselli}, {Moskalenko}, {Murgia},
  {Nakamori}, {Nolan}, {Norris}, {Nuss}, {Ohsugi}, {Omodei}, {Orlando},
  {Ormes}, {Ozaki}, {Paneque}, {Panetta}, {Parent}, {Pelassa}, {Pepe},
  {Pesce-Rollins}, {Piron}, {Porter}, {Rain{\`o}}, {Rando}, {Razzano},
  {Reimer}, {Reimer}, {Reposeur}, {Ritz}, {Rodriguez}, {Romani}, {Roth},
  {Ryde}, {Sadrozinski}, {Sander}, {Saz Parkinson}, {Scargle}, {Sellerholm},
  {Sgr{\`o}}, {Shaw}, {Smith}, {Smith}, {Spandre}, {Spinelli}, {Strickman},
  {Strong}, {Suson}, {Takahashi}, {Tanaka}, {Thayer}, {Thayer}, {Thompson},
  {Tibaldo}, {Tibolla}, {Torres}, {Tosti}, {Tramacere}, {Uchiyama}, {Usher},
  {Vasileiou}, {Vilchez}, {Vitale}, {Waite}, {Wang}, {Winer}, {Wood}, {Ylinen},
  {Ziegler}, \& {Fermi LAT Collaboration}}]{2010ApJ...709L.152A}
---. 2010{\natexlab{b}}, \apjl, 709, L152

\bibitem[{{Acero} {et~al.}(2009){Acero}, {Aharonian}, {Akhperjanian}, {Anton},
  {Barres de Almeida}, {Bazer-Bachi}, {Becherini}, {Behera}, {Bernl{\"o}hr},
  {Bochow}, {Boisson}, {Bolmont}, {Borrel}, {Brucker}, {Brun}, {Brun},
  {B{\"u}hler}, {Bulik}, {B{\"u}sching}, {Boutelier}, {Chadwick},
  {Charbonnier}, {Chaves}, {Cheesebrough}, {Chounet}, {Clapson}, {Coignet},
  {Dalton}, {Daniel}, {Davids}, {Degrange}, {Deil}, {Dickinson},
  {Djannati-Ata{\"i}}, {Domainko}, {Drury}, {Dubois}, {Dubus}, {Dyks}, {Dyrda},
  {Egberts}, {Emmanoulopoulos}, {Espigat}, {Farnier}, {Fegan}, {Feinstein},
  {Fiasson}, {F{\"o}rster}, {Fontaine}, {F{\"u}{\ss}ling}, {Gabici}, {Gallant},
  {G{\'e}rard}, {Gerbig}, {Giebels}, {Glicenstein}, {Gl{\"u}ck}, {Goret},
  {G{\"o}ring}, {Hauser}, {Hauser}, {Heinz}, {Heinzelmann}, {Henri}, {Hermann},
  {Hinton}, {Hoffmann}, {Hofmann}, {Hofverberg}, {Hoppe}, {Horns},
  {Jacholkowska}, {de Jager}, {Jahn}, {Jung}, {Katarzy{\'n}ski}, {Katz},
  {Kaufmann}, {Kerschhaggl}, {Khangulyan}, {Kh{\'e}lifi}, {Keogh}, {Klochkov},
  {Klu{\'z}niak}, {Kneiske}, {Komin}, {Kosack}, {Kossakowski}, {Lamanna},
  {Lenain}, {Lohse}, {Marandon}, {Martineau-Huynh}, {Marcowith}, {Masbou},
  {Maurin}, {McComb}, {Medina}, {M{\'e}hault}, {Moderski}, {Moulin},
  {Naumann-Godo}, {de Naurois}, {Nedbal}, {Nekrassov}, {Nicholas}, {Niemiec},
  {Nolan}, {Ohm}, {Olive}, {Wilhelmi}, {Orford}, {Ostrowski}, {Panter},
  {Arribas}, {Pedaletti}, {Pelletier}, {Petrucci}, {Pita}, {P{\"u}hlhofer},
  {Punch}, {Quirrenbach}, {Raubenheimer}, {Raue}, {Rayner}, {Reimer}, {Renaud},
  {Rieger}, {Ripken}, {Rob}, {Rosier-Lees}, {Rowell}, {Rudak}, {Rulten},
  {Ruppel}, {Sahakian}, {Santangelo}, {Schlickeiser}, {Sch{\"o}ck}, {Schwanke},
  {Schwarzburg}, {Schwemmer}, {Shalchi}, {Sikora}, {Skilton}, {Sol}, {Stawarz},
  {Steenkamp}, {Stegmann}, {Stinzing}, {Superina}, {Szostek}, {Tam},
  {Tavernet}, {Terrier}, {Tibolla}, {Tluczykont}, {van Eldik}, {Vasileiadis},
  {Venter}, {Venter}, {Vialle}, {Vincent}, {Vivier}, {V{\"o}lk}, {Volpe},
  {Wagner}, {Ward}, {Zdziarski}, \& {Zech}}]{2009Sci...326.1080A}
{Acero}, F., {Aharonian}, F., {Akhperjanian}, A.~G., {et~al.} 2009, Science,
  326, 1080

\bibitem[{{Acero} {et~al.}(2015){Acero}, {Ackermann}, {Ajello}, {Albert},
  {Atwood}, {Axelsson}, {Baldini}, {Ballet}, {Barbiellini}, {Bastieri},
  {Belfiore}, {Bellazzini}, {Bissaldi}, {Blandford}, {Bloom}, {Bogart},
  {Bonino}, {Bottacini}, {Bregeon}, {Britto}, {Bruel}, {Buehler}, {Burnett},
  {Buson}, {Caliandro}, {Cameron}, {Caputo}, {Caragiulo}, {Caraveo},
  {Casandjian}, {Cavazzuti}, {Charles}, {Chaves}, {Chekhtman}, {Cheung},
  {Chiang}, {Chiaro}, {Ciprini}, {Claus}, {Cohen-Tanugi}, {Cominsky}, {Conrad},
  {Cutini}, {D'Ammando}, {de Angelis}, {DeKlotz}, {de Palma}, {Desiante},
  {Digel}, {Di Venere}, {Drell}, {Dubois}, {Dumora}, {Favuzzi}, {Fegan},
  {Ferrara}, {Finke}, {Franckowiak}, {Fukazawa}, {Funk}, {Fusco}, {Gargano},
  {Gasparrini}, {Giebels}, {Giglietto}, {Giommi}, {Giordano}, {Giroletti},
  {Glanzman}, {Godfrey}, {Grenier}, {Grondin}, {Grove}, {Guillemot}, {Guiriec},
  {Hadasch}, {Harding}, {Hays}, {Hewitt}, {Hill}, {Horan}, {Iafrate}, {Jogler},
  {J{\'o}hannesson}, {Johnson}, {Johnson}, {Johnson}, {Johnson}, {Kamae},
  {Kataoka}, {Katsuta}, {Kuss}, {La Mura}, {Landriu}, {Larsson}, {Latronico},
  {Lemoine-Goumard}, {Li}, {Li}, {Longo}, {Loparco}, {Lott}, {Lovellette},
  {Lubrano}, {Madejski}, {Massaro}, {Mayer}, {Mazziotta}, {McEnery},
  {Michelson}, {Mirabal}, {Mizuno}, {Moiseev}, {Mongelli}, {Monzani},
  {Morselli}, {Moskalenko}, {Murgia}, {Nuss}, {Ohno}, {Ohsugi}, {Omodei},
  {Orienti}, {Orlando}, {Ormes}, {Paneque}, {Panetta}, {Perkins},
  {Pesce-Rollins}, {Piron}, {Pivato}, {Porter}, {Racusin}, {Rando}, {Razzano},
  {Razzaque}, {Reimer}, {Reimer}, {Reposeur}, {Rochester}, {Romani},
  {Salvetti}, {S{\'a}nchez-Conde}, {Saz Parkinson}, {Schulz}, {Siskind},
  {Smith}, {Spada}, {Spandre}, {Spinelli}, {Stephens}, {Strong}, {Suson},
  {Takahashi}, {Takahashi}, {Tanaka}, {Thayer}, {Thayer}, {Thompson},
  {Tibaldo}, {Tibolla}, {Torres}, {Torresi}, {Tosti}, {Troja}, {Van Klaveren},
  {Vianello}, {Winer}, {Wood}, {Wood}, {Zimmer}, \& {Fermi-LAT
  Collaboration}}]{2015ApJS..218...23A}
{Acero}, F., {Ackermann}, M., {Ajello}, M., {et~al.} 2015, \apjs, 218, 23

\bibitem[{{Ackermann} {et~al.}(2012{\natexlab{a}}){Ackermann}, {Ajello},
  {Allafort}, {Baldini}, {Ballet}, {Bastieri}, {Bechtol}, {Bellazzini},
  {Berenji}, {Bloom}, {Bonamente}, {Borgland}, {Bouvier}, {Bregeon}, {Brigida},
  {Bruel}, {Buehler}, {Buson}, {Caliandro}, {Cameron}, {Caraveo}, {Casandjian},
  {Cecchi}, {Charles}, {Chekhtman}, {Cheung}, {Chiang}, {Cillis}, {Ciprini},
  {Claus}, {Cohen-Tanugi}, {Conrad}, {Cutini}, {de Palma}, {Dermer}, {Digel},
  {Silva}, {Drell}, {Drlica-Wagner}, {Favuzzi}, {Fegan}, {Fortin}, {Fukazawa},
  {Funk}, {Fusco}, {Gargano}, {Gasparrini}, {Germani}, {Giglietto}, {Giordano},
  {Glanzman}, {Godfrey}, {Grenier}, {Guiriec}, {Gustafsson}, {Hadasch},
  {Hayashida}, {Hays}, {Hughes}, {J{\'o}hannesson}, {Johnson}, {Kamae},
  {Katagiri}, {Kataoka}, {Kn{\"o}dlseder}, {Kuss}, {Lande}, {Longo}, {Loparco},
  {Lott}, {Lovellette}, {Lubrano}, {Madejski}, {Martin}, {Mazziotta},
  {McEnery}, {Michelson}, {Mizuno}, {Monte}, {Monzani}, {Morselli},
  {Moskalenko}, {Murgia}, {Nishino}, {Norris}, {Nuss}, {Ohno}, {Ohsugi},
  {Okumura}, {Omodei}, {Orlando}, {Ozaki}, {Parent}, {Persic}, {Pesce-Rollins},
  {Petrosian}, {Pierbattista}, {Piron}, {Pivato}, {Porter}, {Rain{\`o}},
  {Rando}, {Razzano}, {Reimer}, {Reimer}, {Ritz}, {Roth}, {Sbarra}, {Sgr{\`o}},
  {Siskind}, {Spandre}, {Spinelli}, {Stawarz}, {Strong}, {Takahashi}, {Tanaka},
  {Thayer}, {Tibaldo}, {Tinivella}, {Torres}, {Tosti}, {Troja}, {Uchiyama},
  {Vandenbroucke}, {Vianello}, {Vitale}, {Waite}, {Wood}, \&
  {Yang}}]{2012ApJ...755..164A}
{Ackermann}, M., {Ajello}, M., {Allafort}, A., {et~al.} 2012{\natexlab{a}},
  \apj, 755, 164

\bibitem[{{Ackermann} {et~al.}(2012{\natexlab{b}}){Ackermann}, {Ajello},
  {Allafort}, {Baldini}, {Ballet}, {Barbiellini}, {Bastieri}, {Bechtol},
  {Bellazzini}, {Berenji}, {Bloom}, {Bonamente}, {Borgland}, {Bregeon},
  {Brigida}, {Bruel}, {Buehler}, {Buson}, {Caliandro}, {Cameron}, {Caraveo},
  {Casandjian}, {Cavazzuti}, {Cecchi}, {Charles}, {Chekhtman}, {Cheung},
  {Chiang}, {Ciprini}, {Claus}, {Cohen-Tanugi}, {Conrad}, {Cutini},
  {D'Ammando}, {de Angelis}, {de Palma}, {Dermer}, {Silva}, {Drell},
  {Drlica-Wagner}, {Enoto}, {Favuzzi}, {Fegan}, {Ferrara}, {Fortin},
  {Fukazawa}, {Fusco}, {Gargano}, {Gasparrini}, {Gehrels}, {Germani},
  {Giglietto}, {Giommi}, {Giordano}, {Giroletti}, {Godfrey}, {Grove},
  {Guiriec}, {Hadasch}, {Hayashida}, {Hays}, {Hughes}, {J{\'o}hannesson},
  {Johnson}, {Kamae}, {Katagiri}, {Kataoka}, {Kn{\"o}dlseder}, {Kuss}, {Lande},
  {Llena Garde}, {Longo}, {Loparco}, {Lott}, {Lovellette}, {Lubrano},
  {Madejski}, {Mazziotta}, {Michelson}, {Mizuno}, {Monte}, {Monzani},
  {Morselli}, {Moskalenko}, {Murgia}, {Nishino}, {Norris}, {Nuss}, {Ohno},
  {Ohsugi}, {Okumura}, {Orlando}, {Ozaki}, {Paneque}, {Pesce-Rollins},
  {Pierbattista}, {Piron}, {Pivato}, {Porter}, {Rain{\`o}}, {Rando}, {Razzano},
  {Reimer}, {Reimer}, {Ritz}, {Roth}, {Sanchez}, {Sbarra}, {Sgr{\`o}},
  {Siskind}, {Spandre}, {Spinelli}, {Stawarz}, {Strong}, {Takahashi},
  {Takahashi}, {Tanaka}, {Thayer}, {Thompson}, {Tibaldo}, {Tinivella},
  {Torres}, {Tosti}, {Troja}, {Uchiyama}, {Usher}, {Vandenbroucke},
  {Vasileiou}, {Vianello}, {Vitale}, {Waite}, {Winer}, {Wood}, {Wood}, {Yang},
  \& {Zimmer}}]{2012ApJ...747..104A}
---. 2012{\natexlab{b}}, \apj, 747, 104

\bibitem[{{Ajello} {et~al.}(2008){Ajello}, {Rau}, {Greiner}, {Kanbach},
  {Salvato}, {Strong}, {Barthelmy}, {Gehrels}, {Markwardt}, \&
  {Tueller}}]{2008ApJ...673...96A}
{Ajello}, M., {Rau}, A., {Greiner}, J., {et~al.} 2008, \apj, 673, 96

\bibitem[{{Caballero-Garcia} {et~al.}(2012){Caballero-Garcia}, {Papadakis},
  {Nicastro}, \& {Ajello}}]{2012A&A...537A..87C}
{Caballero-Garcia}, M.~D., {Papadakis}, I.~E., {Nicastro}, F., \& {Ajello}, M.
  2012, \aap, 537, A87

\bibitem[{{Cerutti} {et~al.}(2011){Cerutti}, {Dubus}, {Malzac}, {Szostek},
  {Belmont}, {Zdziarski}, \& {Henri}}]{2011A&A...529A.120C}
{Cerutti}, B., {Dubus}, G., {Malzac}, J., {et~al.} 2011, \aap, 529, A120

\bibitem[{{Corbel} {et~al.}(2012){Corbel}, {Dubus}, {Tomsick}, {Szostek},
  {Corbet}, {Miller-Jones}, {Richards}, {Pooley}, {Trushkin}, {Dubois}, {Hill},
  {Kerr}, {Max-Moerbeck}, {Readhead}, {Bodaghee}, {Tudose}, {Parent}, {Wilms},
  \& {Pottschmidt}}]{2012MNRAS.421.2947C}
{Corbel}, S., {Dubus}, G., {Tomsick}, J.~A., {et~al.} 2012, \mnras, 421, 2947

\bibitem[{{Done} {et~al.}(2007){Done}, {Gierli{\'n}ski}, \&
  {Kubota}}]{2007A&ARv..15....1D}
{Done}, C., {Gierli{\'n}ski}, M., \& {Kubota}, A. 2007, \aapr, 15, 1

\bibitem[{{Done} {et~al.}(2003){Done}, {Madejski}, {{\.Z}ycki}, \&
  {Greenhill}}]{2003ApJ...588..763D}
{Done}, C., {Madejski}, G.~M., {{\.Z}ycki}, P.~T., \& {Greenhill}, L.~J. 2003,
  \apj, 588, 763

\bibitem[{{Eichmann} \& {Becker Tjus}(2016)}]{2016ApJ...821...87E}
{Eichmann}, B., \& {Becker Tjus}, J. 2016, \apj, 821, 87

\bibitem[{{Elmouttie} {et~al.}(1998){Elmouttie}, {Haynes}, {Jones}, {Sadler},
  \& {Ehle}}]{1998MNRAS.297.1202E}
{Elmouttie}, M., {Haynes}, R.~F., {Jones}, K.~L., {Sadler}, E.~M., \& {Ehle},
  M. 1998, \mnras, 297, 1202

\bibitem[{{Fasano} \& {Franceschini}(1987)}]{1987MNRAS.225..155F}
{Fasano}, G., \& {Franceschini}, A. 1987, \mnras, 225, 155

\bibitem[{{Fender} {et~al.}(2004){Fender}, {Belloni}, \&
  {Gallo}}]{2004MNRAS.355.1105F}
{Fender}, R.~P., {Belloni}, T.~M., \& {Gallo}, E. 2004, \mnras, 355, 1105

\bibitem[{{Gallimore} {et~al.}(2006){Gallimore}, {Axon}, {O'Dea}, {Baum}, \&
  {Pedlar}}]{2006AJ....132..546G}
{Gallimore}, J.~F., {Axon}, D.~J., {O'Dea}, C.~P., {Baum}, S.~A., \& {Pedlar},
  A. 2006, \aj, 132, 546

\bibitem[{{Greenhill} {et~al.}(1997){Greenhill}, {Moran}, \&
  {Herrnstein}}]{1997ApJ...481L..23G}
{Greenhill}, L.~J., {Moran}, J.~M., \& {Herrnstein}, J.~R. 1997, \apjl, 481,
  L23

\bibitem[{{Griffin} {et~al.}(2016){Griffin}, {Dai}, \&
  {Thompson}}]{2016ApJ...823L..17G}
{Griffin}, R.~D., {Dai}, X., \& {Thompson}, T.~A. 2016, \apjl, 823, L17

\bibitem[{{Hayashida} {et~al.}(2013){Hayashida}, {Stawarz}, {Cheung},
  {Bechtol}, {Madejski}, {Ajello}, {Massaro}, {Moskalenko}, {Strong}, \&
  {Tibaldo}}]{2013ApJ...779..131H}
{Hayashida}, M., {Stawarz}, {\L}., {Cheung}, C.~C., {et~al.} 2013, \apj, 779,
  131

\bibitem[{{Kennicutt}(1998)}]{1998ApJ...498..541K}
{Kennicutt}, Jr., R.~C. 1998, \apj, 498, 541

\bibitem[{{Kimura} {et~al.}(2016){Kimura}, {Toma}, {Suzuki}, \&
  {Inutsuka}}]{2016ApJ...822...88K}
{Kimura}, S.~S., {Toma}, K., {Suzuki}, T.~K., \& {Inutsuka}, S.-i. 2016, \apj,
  822, 88

\bibitem[{{Krimm} {et~al.}(2013){Krimm}, {Holland}, {Corbet}, {Pearlman},
  {Romano}, {Kennea}, {Bloom}, {Barthelmy}, {Baumgartner}, {Cummings},
  {Gehrels}, {Lien}, {Markwardt}, {Palmer}, {Sakamoto}, {Stamatikos}, \&
  {Ukwatta}}]{2013ApJS..209...14K}
{Krimm}, H.~A., {Holland}, S.~T., {Corbet}, R.~H.~D., {et~al.} 2013, \apjs,
  209, 14

\bibitem[{{Lacki} {et~al.}(2011){Lacki}, {Thompson}, {Quataert}, {Loeb}, \&
  {Waxman}}]{2011ApJ...734..107L}
{Lacki}, B.~C., {Thompson}, T.~A., {Quataert}, E., {Loeb}, A., \& {Waxman}, E.
  2011, \apj, 734, 107

\bibitem[{{Lamastra} {et~al.}(2016){Lamastra}, {Fiore}, {Guetta}, {Antonelli},
  {Colafrancesco}, {Menci}, {Puccetti}, {Stamerra}, \&
  {Zappacosta}}]{2016A&A...596A..68L}
{Lamastra}, A., {Fiore}, F., {Guetta}, D., {et~al.} 2016, \aap, 596, A68

\bibitem[{{Lenain} {et~al.}(2010){Lenain}, {Ricci}, {T{\"u}rler}, {Dorner}, \&
  {Walter}}]{2010A&A...524A..72L}
{Lenain}, J.-P., {Ricci}, C., {T{\"u}rler}, M., {Dorner}, D., \& {Walter}, R.
  2010, \aap, 524, A72

\bibitem[{{Lenc} \& {Tingay}(2009)}]{2009AJ....137..537L}
{Lenc}, E., \& {Tingay}, S.~J. 2009, \aj, 137, 537

\bibitem[{{Madejski} {et~al.}(2000){Madejski}, {{\.Z}ycki}, {Done}, {Valinia},
  {Blanco}, {Rothschild}, \& {Turek}}]{2000ApJ...535L..87M}
{Madejski}, G., {{\.Z}ycki}, P., {Done}, C., {et~al.} 2000, \apjl, 535, L87

\bibitem[{{Malzac}(2016)}]{2016AN....337..391M}
{Malzac}, J. 2016, Astronomische Nachrichten, 337, 391

\bibitem[{{Marscher} {et~al.}(2002){Marscher}, {Jorstad}, {G{\'o}mez}, {Aller},
  {Ter{\"a}sranta}, {Lister}, \& {Stirling}}]{2002Natur.417..625M}
{Marscher}, A.~P., {Jorstad}, S.~G., {G{\'o}mez}, J.-L., {et~al.} 2002, \nat,
  417, 625

\bibitem[{{Nied{\'z}wiecki} {et~al.}(2013){Nied{\'z}wiecki}, {Xie}, \&
  {Stepnik}}]{2013MNRAS.432.1576N}
{Nied{\'z}wiecki}, A., {Xie}, F.-G., \& {Stepnik}, A. 2013, \mnras, 432, 1576

\bibitem[{{Ohm}(2016)}]{2016CRPhy..17..585O}
{Ohm}, S. 2016, Comptes Rendus Physique, 17, 585

\bibitem[{{Poutanen} \& {Veledina}(2014)}]{2014SSRv..183...61P}
{Poutanen}, J., \& {Veledina}, A. 2014, \ssr, 183, 61

\bibitem[{{Press} {et~al.}(1992){Press}, {Teukolsky}, {Vetterling}, \&
  {Flannery}}]{1992nrfa.book.....P}
{Press}, W.~H., {Teukolsky}, S.~A., {Vetterling}, W.~T., \& {Flannery}, B.~P.
  1992, {Numerical recipes in FORTRAN. The art of scientific computing}

\bibitem[{{Puccetti} {et~al.}(2014){Puccetti}, {Comastri}, {Fiore},
  {Ar{\'e}valo}, {Risaliti}, {Bauer}, {Brandt}, {Stern}, {Harrison},
  {Alexander}, {Boggs}, {Christensen}, {Craig}, {Gandhi}, {Hailey}, {Koss},
  {Lansbury}, {Luo}, {Madejski}, {Matt}, {Walton}, \&
  {Zhang}}]{2014ApJ...793...26P}
{Puccetti}, S., {Comastri}, A., {Fiore}, F., {et~al.} 2014, \apj, 793, 26

\bibitem[{{Rojas-Bravo} \& {Araya}(2016)}]{2016MNRAS.463.1068R}
{Rojas-Bravo}, C., \& {Araya}, M. 2016, \mnras, 463, 1068

\bibitem[{{Shakura} \& {Sunyaev}(1973)}]{1973A&A....24..337S}
{Shakura}, N.~I., \& {Sunyaev}, R.~A. 1973, \aap, 24, 337

\bibitem[{{Tang} {et~al.}(2014){Tang}, {Wang}, \& {Tam}}]{2014ApJ...794...26T}
{Tang}, Q.-W., {Wang}, X.-Y., \& {Tam}, P.-H.~T. 2014, \apj, 794, 26

\bibitem[{{VERITAS Collaboration} {et~al.}(2009){VERITAS Collaboration},
  {Acciari}, {Aliu}, {Arlen}, {Aune}, {Bautista}, {Beilicke}, {Benbow},
  {Boltuch}, {Bradbury}, {Buckley}, {Bugaev}, {Byrum}, {Cannon}, {Celik},
  {Cesarini}, {Chow}, {Ciupik}, {Cogan}, {Colin}, {Cui}, {Dickherber}, {Duke},
  {Fegan}, {Finley}, {Finnegan}, {Fortin}, {Fortson}, {Furniss}, {Galante},
  {Gall}, {Gibbs}, {Gillanders}, {Godambe}, {Grube}, {Guenette}, {Gyuk},
  {Hanna}, {Holder}, {Horan}, {Hui}, {Humensky}, {Imran}, {Kaaret}, {Karlsson},
  {Kertzman}, {Kieda}, {Kildea}, {Konopelko}, {Krawczynski}, {Krennrich},
  {Lang}, {Lebohec}, {Maier}, {McArthur}, {McCann}, {McCutcheon}, {Millis},
  {Moriarty}, {Mukherjee}, {Nagai}, {Ong}, {Otte}, {Pandel}, {Perkins},
  {Pizlo}, {Pohl}, {Quinn}, {Ragan}, {Reyes}, {Reynolds}, {Roache}, {Rose},
  {Schroedter}, {Sembroski}, {Smith}, {Steele}, {Swordy}, {Theiling},
  {Thibadeau}, {Varlotta}, {Vassiliev}, {Vincent}, {Wagner}, {Wakely}, {Ward},
  {Weekes}, {Weinstein}, {Weisgarber}, {Williams}, {Wissel}, {Wood}, \&
  {Zitzer}}]{2009Natur.462..770V}
{VERITAS Collaboration}, {Acciari}, V.~A., {Aliu}, E., {et~al.} 2009, \nat,
  462, 770

\bibitem[{{Wojaczy{\'n}ski} {et~al.}(2015){Wojaczy{\'n}ski}, {Nied{\'z}wiecki},
  {Xie}, \& {Szanecki}}]{2015A&A...584A..20W}
{Wojaczy{\'n}ski}, R., {Nied{\'z}wiecki}, A., {Xie}, F.-G., \& {Szanecki}, M.
  2015, \aap, 584, A20

\bibitem[{{Yoast-Hull} {et~al.}(2014){Yoast-Hull}, {Gallagher}, {Zweibel}, \&
  {Everett}}]{2014ApJ...780..137Y}
{Yoast-Hull}, T.~M., {Gallagher}, III, J.~S., {Zweibel}, E.~G., \& {Everett},
  J.~E. 2014, \apj, 780, 137

\end{thebibliography}

\end{document}